\documentclass[aip,jcp,graphicx,longbibliography,reprint,citeautoscript]{revtex4-1}

\draft 

\usepackage{amssymb}
\usepackage{mathtools}
\usepackage[version=4]{mhchem}
\usepackage{graphicx}
\usepackage{hyperref}
\usepackage{color}
\usepackage{afterpage}
\usepackage{bm}
\usepackage{enumitem}
\usepackage{wrapfig}
\usepackage{subfigure}

\newcommand{\NLi}{N_{\text{Li}}}
\newcommand{\NSi}{N_{\text{Si}}}

\newcommand{\C}{\mathbb{C}}
\newcommand{\N}{\mathbb{N}}
\newcommand{\R}{\mathbb{R}}
\newcommand{\Z}{\mathbb{Z}}
\newcommand{\Xc}{\mathcal{X}}
\newcommand{\Qc}{\mathcal{Q}}

\newcommand{\Lb}{\mathbf{L}}
\newcommand{\Gc}{\mathcal{G}}
\newcommand{\Fc}{\mathcal{F}}

\DeclareMathOperator*{\arginf}{arg\,inf}

\DeclarePairedDelimiter{\abs}{\lvert}{\rvert} 
\DeclarePairedDelimiter{\norm}{\lVert}{\rVert}

\begin{document}

\title{Wavelet Scattering Networks for Atomistic Systems with Extrapolation of Material Properties} 



\author{Paul Sinz}
\thanks{These authors contributed equally to the work.}
\affiliation{Department of Computational Mathematics, Science and Engineering, Michigan
State University, East Lansing, Michigan 48824-1226, USA}

\author{Michael W. Swift}
\thanks{These authors contributed equally to the work.}
\affiliation{Department of Chemical Engineering and Materials Science, Michigan
State University, East Lansing, Michigan 48824-1226, USA}

\author{Xavier Brumwell}
\affiliation{Department of Computational Mathematics, Science and Engineering, Michigan
State University, East Lansing, Michigan 48824-1226, USA}

\author{Jialin Liu}
\affiliation{Department of Chemical Engineering and Materials Science, Michigan
State University, East Lansing, Michigan 48824-1226, USA}

\author{Kwang Jin Kim}
\affiliation{Department of Chemical Engineering and Materials Science, Michigan
State University, East Lansing, Michigan 48824-1226, USA}

\author{Yue Qi}
\email{yueqi@egr.msu.edu}
\affiliation{Department of Chemical Engineering and Materials Science, Michigan
State University, East Lansing, Michigan 48824-1226, USA}

\author{Matthew Hirn}
\email{mhirn@msu.edu}
\affiliation{Department of Computational Mathematics, Science and Engineering, Michigan
State University, East Lansing, Michigan 48824-1226, USA}
\affiliation{Department of Mathematics, Michigan State University, East Lansing, Michigan 48824-1226, USA}
\affiliation{Center for Quantum Computing, Science and Engineering, Michigan State University, East Lansing, Michigan 48824-1226, USA}


\date{\today}

\begin{abstract}
The dream of machine learning in materials science is for a model to learn the underlying physics of an atomic system, allowing it to move beyond interpolation of the training set to the prediction of properties that were not present in the original training data. In addition to advances in machine learning architectures and training techniques, achieving this ambitious goal requires a method to convert a 3D atomic system into a feature representation that preserves rotational and translational symmetry, smoothness under small perturbations, and invariance under re-ordering.  The atomic orbital wavelet scattering transform preserves these symmetries by construction, and has achieved great success as a featurization method for machine learning energy prediction. Both in small molecules and in the bulk amorphous \ce{Li_{\alpha}Si} system, machine learning models using wavelet scattering coefficients as features have demonstrated a comparable accuracy to Density Functional Theory at a small fraction of the computational cost. In this work, we test the generalizability of our \ce{Li_{\alpha}Si} energy predictor to properties that were not included in the training set, such as elastic constants and migration barriers. We demonstrate that statistical feature selection methods can reduce over-fitting and lead to remarkable accuracy in these extrapolation tasks.
\end{abstract}


\maketitle 

\section{Introduction}

Machine learning (ML) is a powerful tool in chemical physics~\cite{butler:mlMoleculesMaterials2018, Noe:MLMolecularSimulation2019}.  Both kernel-based~\cite{Muller:kernelIntro2001, bartok:gaussAppPot2010, rupp:coulombMatrix2012, PhysRevLett.109.059801, bartok:repChemEnviron2013, montavon:mlMolProp2013, De:moleculesAcrossAlchemicalSpace2016, Shapeev2015-MTP, chmiela:MLconservativeFF2017, burke:bypassingKohnSham2017, chimela:ML-MD2018, Bereau:phys-basedPotentials2018} and neural-network-based~\cite{goodfellow:deeplearningBook2016, Behler:NNPotEnergy2007, Behler:atomisticSimulations2010, schutt:qcDeepTensor2016, schutt:molecuLeNet2017, smith:NNpotential2017, gilmer:MPNNQC2017, hy:molecularPropsCovariantNN2018, zhang:deepPotentialMD2018, smidt:tensorFieldNetworks2018} learning algorithms have found success predicting physical properties such as energies, forces, and potential energy surfaces starting from atomic coordinates.  ML models have been used for molecular dynamics (MD)~\cite{Artrith:LiSiML2018,chimela:ML-MD2018,PhysRevLett.114.096405}, prediction of free energy surfaces~\cite{onat:implantNNLiSi2018,Stecher:FreeEnergySurfaceGPR2014,Mones:FreeEnergySurfaceGPR16,Schneider:StochasticNN-FES2017}, and generation of thermodynamic ensembles~\cite{Artrith:LiSiML2018,Noe:BoltzmannGenerators2019} on systems for which they have been trained.  Much as ML models have revolutionized fields like computer vision~\cite{Voulodimos:ComputerVisionReview2018}, automated content generation~\cite{gatt2018survey}, and natural language processing~\cite{Hirschberg:NLP2015}, an ML model could in principle predict physical properties of broad classes of atomic systems with accuracy competitive with the best current methods at a small fraction of the computational cost. However, such an ML model has not yet been developed and many obstacles still remain before general atomistic ML models can be competitive with existing quantum chemistry methods.  The most fundamental obstacle is perhaps the generalizability problem (also referred to as transferability).  Quantum chemical methods such as density functional theory (DFT) are predictive because they work from first principles---physical properties emerge from solutions to equations that describe underlying physics, allowing these techniques to work on systems that have never been studied before.  In contrast, machine learning is at its heart a fitting technique.  The model knows nothing about the physical equations and its predictions are statistical inferences based on the training data.  This raises the question: are quantum chemistry models based on machine learning inherently limited to interpolation of the training data?  Or can a machine learning model ``learn the underlying physics'' and provide new insights beyond the training data?  If we want to answer a question with an ML model, will it always be necessary to compute the answer to the same question beforehand in thousands or tens of thousands of similar cases?  Or could ML truly teach us something new?  The hope that it can is not entirely without foundation.  Machine learning models have shown the ability to capture and generalize abstract patterns within training data.  For example, a neural net trained to translate languages recently achieved ``zero-shot translation'': translation between a pair of languages for which it had no dictionary without going through an intermediate language~\cite{johnson2016googles}.  This suggests that learning of underlying meaning from diverse examples may be possible.  

Of course the full answer to this question is beyond the scope of the present work.  We look at a specific example of the generalizability problem: energy prediction for amorphous \ce{Li_{\alpha}Si}. Due to its potential for developing high energy density lithium-ion batteries, this system has recently been studied by a variety of machine learning methods.  Onat \emph{et al.} generated an implanted neural network potential for this system~\cite{onat:implantNNLiSi2018}.  Artrith \emph{et al.} developed a machine-learning potential that enabled ensemble generation and molecular dynamics for \ce{Li_{\alpha}Si}~\cite{Artrith:LiSiML2018}.  Brumwell \emph{et al.} created a three-dimensional ML model, similar to a convolutional neural network, with a physically motivated filter based on the wavelet scattering transform~\cite{mallat:scattering2012}, and were able to achieve chemical accuracy in energy prediction for this class of structures.~\cite{brumwell2018steerable}  The wavelet formulation has a number of advantages.  The inclusion of wavelet dilations and second-order scattering coefficients makes the wavelet scattering approach inherently multiscale, including length scales on the order of the support of the smallest wavelet up to the length scale of the largest unit cell in the \ce{Li_{\alpha}Si} database. This allows the wavelet scattering transform to capture a broad class of physical interactions, making it more general (see section~\ref{subsec:ML_algorithm} for details on the multiscale nature).  The fact that the wavelets themselves are based on atomic orbitals may allow them to more naturally capture interactions arising from electronic bonds (though without sacrificing generality since the wavelet frame is overcomplete) and may help with generalizability of the model. Other methods which sum up atomic energies based on their local environments (e.g. Refs.~\citenum{bartok:repChemEnviron2013, Artrith:LiSiML2018, onat:implantNNLiSi2018, imbalzano:atomicFingerprints2018, drautz:ACE2019}) can similarly separate length scales by setting the cutoff to the length scale of the unit cell and selecting a variety of radial functions with support along the intermediate length scales down to the distance between nearest-neighbor atoms. These methods provide feature descriptions which separate length scales, but must be used in an appropriate training or energy fitting framework in order to nonlinearly couple these scales. In contrast, the wavelet scattering coefficients used here include the nonlinear coupling in the feature set.

In this work, we build on and improve the wavelet-based model~\cite{hirn:waveletScatQuantum2016, eickenberg:3DSolidHarmonicScat2017, eickenberg:scatMoleculesJCP2018, brumwell2018steerable}, achieving similar accuracy to ~\citet{onat:implantNNLiSi2018} and ~\citet{Artrith:LiSiML2018} but using a simpler architecture with fewer parameters than the neural network approaches and with fewer features than any of the previous models including~ \citet{brumwell2018steerable} More importantly, we test the model on extrapolation tasks, predicting physical properties that were not present in the training set. We perform detailed analysis on how to balance under-fitting and over-fitting in order to achieve high generalizability. The three tasks are predicting migration barriers based on transition state theory, energies of amorphous systems significantly larger than the training set systems, and elastic properties based on deformations of amorphous \ce{Li_{\alpha}Si}.  In each of these extrapolation tasks, we find that the model is able to achieve reasonable accuracy, and in some cases does quite well, thus providing evidence of the model's ability to generalize to new types of systems and tasks. 

The remainder of this paper is organized as follows. In Section \ref{sec:methods} we present the methods used in this work, including the data generation process and descriptions of the machine learning algorithms. Numerical results are presented and discussed in Section \ref{sec:results}, and Section \ref{sec: conclusion} contains a short conclusion. Appendices and Supplemental Material provide additional details on certain aspects of the paper.

\section{Methods}
\label{sec:methods}

Methods consist of data generation for training (Section \ref{subsec:training_data_generation}) as well as data generation for testing on extrapolation tasks (Section \ref{sec: data gen for extrapolation}). Algorithms used for training the machine learned models are discussed in Section \ref{sec: linear regression and model fitting}, whereas the wavelet scattering representation of an atomic state is described in Section \ref{subsec:ML_algorithm}. Appendix \ref{sec: fast wavelet scattering computations} explains how to compute such features efficiently.

\subsection{Training data generation}
\label{subsec:training_data_generation}

\begin{figure}
    \centering
    \includegraphics[width=\columnwidth]{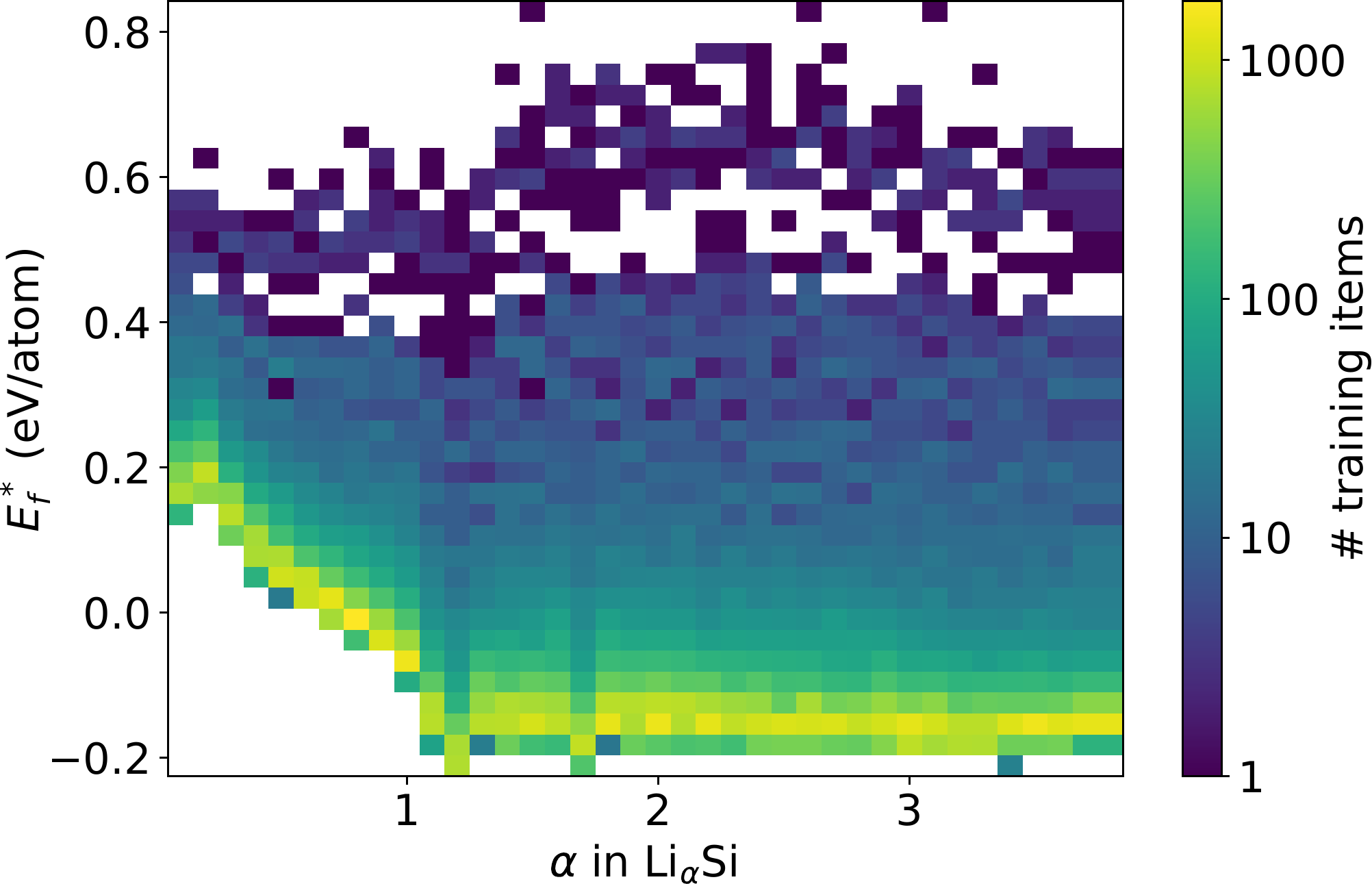}
    \caption{Histogram of training set energies versus concentration $\alpha$ in \ce{Li_{\alpha}Si}.  Color indicates the the number of training items in each bin on a logarithmic scale. }
    \label{fig:dataset}
\end{figure}

The training, validation, and interpolation testing data for the machine-learned model consists of amorphous \ce{Li_{\alpha}Si} structures labeled by formation energies calculated using Density Functional Theory (DFT). These structures are in cubic boxes under periodic boundary conditions containing from 55 to 100 atoms, with lithium-to-silicon ratio $\alpha$ ranging from 0.1 to 3.75.  Initial disordered structures are generated by evolving random structures under ReaxFF~\cite{Senftle:ReaxFF2016} molecular dynamics (MD) at 2500K for 10 ps, and ten different disordered structures are randomly picked from the MD trajectory for each of the 37 chosen concentrations. The accuracy of the force field used to obtain the initial amorphous structure is not important, due to the following DFT calculations. In particular, each structure is fully relaxed at constant volume using DFT. The structures and formation energies along the relaxation paths make up the amorphous dataset used in this work, which contains a total of 90,252 structures.  A histogram of the quantity of these structures by energy and concentration is shown in Figure~\ref{fig:dataset}.  We note that the structures are heavily concentrated near the endpoint of the relaxation, so we expect the resulting model to do better on near-equilibrium amorphous structures.  This is desirable because the low-energy structures are more likely to arise in realistic simulations. We also calculate voltages versus Li/\ce{Li+}~\cite{Yuan20,Swift19,Liang17} and radial distribution functions for Si-Si, Li-Si, and Li-Li pairs, and find good agreement with similar data sets from the literature~\cite{Artrith:LiSiML2018,Hatchard04} (see the \hyperref[sec: supplementary material]{Supplementary Material}).  This confirms that our amorphous structures are physically realistic.

Formation energies and relaxations were performed in the Vienna Ab initio Simulation Package (VASP)~\cite{Kresse96} using the Projector-Augmented Wave method~\cite{Blochl94} and the PBE exchange-correlation functional~\cite{Perdew96} with a plane-wave energy cutoff of 500 eV.  The Brillouin zone was sampled using the Gamma point only during relaxation.  After relaxation, the energies along each relaxation path were corrected for $k$-point sampling errors by calculating the energy of each fully relaxed structure using a $3 \times 3 \times 3$ Gamma-centered grid and applying the resulting constant shift to the rest of the structures in the relaxation path.  The mean absolute $k$-point sampling correction was 27 meV/atom.  The total formation energy of a structure with $\NLi$ lithium atoms and $\NSi$ silicon atoms is defined based on DFT total energies:
\begin{equation*}
    E_f (\ce{Li}_{\NLi} \ce{Si}_{\NSi}) = E_\text{tot}(\ce{Li}_{\NLi} \ce{Si}_{\NSi}) - \NLi E(\ce{Li}) - \NSi E(\ce{Si}) \, ,
\end{equation*}
where $E_{\text{tot}}(\ce{Li}_{\NLi} \ce{Si}_{\NSi})$ is the total energy of the system, and $E(\ce{Li})$ and $E(\ce{Si})$ are the DFT total energy per atom of elemental lithium and silicon, respectively.  The structure $\ce{Li}_{\NLi} \ce{Si}_{\NSi}$ has reduced formula \ce{Li_{\alpha}Si} with $\alpha = \NLi / \NSi$ and per-atom formation energy
\begin{equation}
    E_f^*(\ce{Li_{\alpha} Si}) = E_f(\ce{Li}_{\NLi} \ce{Si}_{\NSi}) / (\NLi + \NSi) \, .
\end{equation}
The per-atom formation energy is the quantity of interest for machine learning. Notice, though, it includes the terms $\NLi E(\ce{Li})$ and $\NSi E(\ce{Si})$ which require no additional quantum mechanical calculations beyond the one-time cost of computing $E(\ce{Li})$ and $E(\ce{Si})$. The difficulty is in computing $E_{\text{tot}} (\ce{Li}_{\NLi} \ce{Si}_{\NSi})$, which requires a costly DFT calculation for each new state. When fitting our machine learned models, we regress the per-atom total energy, defined as:
\begin{equation*}
    E_{\text{tot}}^* (\ce{Li_{\alpha} Si}) = E_{\text{tot}} (\ce{Li}_{\NLi} \ce{Si}_{\NSi}) / (\NLi + \NSi)
    \end{equation*}
or 
    \begin{equation*}
   E_{\text{tot}}^* (\ce{Li_{\alpha} Si}) = E_f^*(\ce{Li_{\alpha} Si}) + \frac{\alpha}{1+\alpha}E(\ce{Li}) + \frac{1}{1+\alpha}E(\ce{Si})\, .
\end{equation*}

Even though it is simple to convert total energies into per-atom total energies, we regress the latter since per atom energies remove the effect of varying unit cell sizes and the number of atoms per unit cell on the total energy. Since we use the squared loss as our measure of error when training, regressing total energies would bias the models towards systems containing larger numbers of atoms since the total energy scales with the number of atoms. 

\subsection{Data generation for extrapolation tests}
\label{sec: data gen for extrapolation}

In order to test the machine learning model's generalizability to extrapolation tasks, additional DFT data is required to compare with the results of the machine learning model.  We test three different extrapolation tasks: prediction of migration barriers, energy prediction for systems with larger unit cells, and prediction of elastic properties.

Diffusion barriers cannot be defined uniquely in amorphous structures due to the lack of order.  Rather, paths which move an atom from one favorable coordination environment to another through a relatively unfavorable environment are abundant.  An endpoint for such a pathway was found by locating void spaces in the amorphous structure through Voronoi analysis and inserting a test lithium atom at each void to find the most energetically favorable position.  Nearby lithium atoms to this void were subsequently identified, and the minimum-energy path (MEP) for each lithium to travel to the void was calculated using the nudged elastic band (NEB) method~\cite{Johnsson98}.  The NEB images along 6 calculated MEPs (for a total of 50 image structures) were used as testing data for this extrapolation task.  The primary quantity of interest is the migration barrier, i.e., the difference between the lowest-energy and highest-energy points along the MEP.

Large structure testing data was generated by two methods: independently relaxing larger AIMD-generated structures (the ``from-scratch'' method) or tiling structures from the data set, randomly perturbing all atomic positions by 0.1 \AA, and performing a single-point calculation (the ``tiled'' approach).  The testing data consists of 37 from-scratch structures, 40 $2\times2\times2$ tiled structures, and 108 $2\times1\times1$ tiled structures.  

Finally, elastic property testing data was generated by applying hydrostatic strain from $-9\%$ to $+9\%$ on fully relaxed structures at each concentration.  The bulk modulus $K$ is calculated by fitting data near the minimum to the equation~\cite{Birch:modulus1947}:
\begin{equation*}
    K = V_0 \frac{\partial^2 E}{\partial V^2}
\end{equation*}
where $V$ is the volume, $V_0$ is the equilibrium volume, and $E$ is the energy.  In total a bulk modulus value is calculated at each of the 37 concentrations, based on a total of 333 structures under hydrostatic strain.

\subsection{Linear regression and model fitting}
\label{sec: linear regression and model fitting}

The purpose of using machine learning is to reduce the computational burden relative to quantum mechanical calculations while maintaining accurate predictions. To accomplish this task, we need to derive features from our data and recombine them in some meaningful way. We use a linear model over a predefined set of features, which consist of nonlinear, multiscale maps of the original atomic state.  Using a linear model over a universal feature set allows us to leverage several well-studied techniques in regression, regularization, and statistical learning theory that increases the accuracy, stability, and generalizability of the model. 

Let $x = \{ (Z_k, R_k) \in \N \times \R^3 \}_{k=1}^{N_x}$ denote the list of atoms in the unit cell of a $\ce{Li}_{\alpha} \ce{Si}$ system. The value $Z_k \in \N$ denotes the protonic charge of the atom, i.e., $Z_k = 3$ for lithium and $Z_k = 14$ for silicon, and $R_k \in \R^3$ denotes the position of the atom in the simulation cell. The quantity $N_x = \NLi + \NSi$ is the total number of atoms in the unit cell. Let $\Phi (x) \in \R^d$ be a $d$-dimensional representation of the state $x$, which is described in detail in Section \ref{subsec:ML_algorithm}. A linear regression with weights $w = (w_{\gamma})_{\gamma = 0}^d \in \R^{d+1}$ of the per-atom total energy $E_{\text{tot}}^{\ast} (x)$ over the representation $\Phi (x) = (\phi_{\gamma} (x))_{\gamma = 1}^d$ computes
\begin{equation} \label{eqn: ml energy model}
    \widetilde{E}_{\text{tot}}^{\ast} (x; w) = \widetilde{E} (x; w) =  w_0 + \sum_{\gamma = 1}^d w_{\gamma} \phi_{\gamma} (x) \, ,
\end{equation}
where $w_0$ is a bias term and the coordinates $\phi_{\gamma} (x)$ of $\Phi (x)$ are weighted with the scalars $w_{\gamma}$. We regularize the regression by selecting a parsimonious model via a sparsity constraint on the weights $w$, 
\begin{equation*}
    ||w||_0 = \# \{ w_{\gamma} \neq 0 : 0 \leq \gamma \leq d \} \leq M \, ,
\end{equation*} 
for some hyper-parameter $M$ that determines the number of nonzero weights. 

The weights $(w_{\gamma})_{\gamma = 0}^d$ and the hyper-parameter $M$ are solved for using the DFT generated training data. Let $\Xc_t = \{ (x_i, E(x_i)) \}_{i=1}^{n_t}$ denote a training set consisting of $\ce{Li}_{\alpha} \ce{Si}$ states $x_i$ and their DFT generated per-atom total energies $E(x_i) = E_{\text{tot}}^{\ast} (x_i)$; denote by $\Xc_v = \{ (x_i', E(x_i') \}_{i=1}^{n_v}$ another such set, also consisting of $\ce{Li}_{\alpha} \ce{Si}$ states and their associated per-atom total energies, non-overlapping with the training set, which we use as the validation set. For each $M$ up to some maximum value, $1 \leq M \leq M_{\max}$, we compute weights $w^M = (w_{\gamma}^M)_{\gamma=0}^d$ by solving the following:
\begin{equation} \label{eqn: M-sparse weights}
    w^M = \arginf_{w \in \R^{d+1}} \big[ L (w, \Xc_t) : \| w \|_0 \leq M \big] \, ,
\end{equation}
where $L (w, \Xc_t)$ is the mean squared loss function with respect to the training set $\Xc_t$: 
\begin{equation*}
    L(w, \Xc_t) = \frac{1}{n_t} \sum_{i=1}^{n_t} |E(x_i) - \widetilde{E}(x_i; w)|^2 \, .
\end{equation*}
As $M$ increases the model $\widetilde{E}(x; w^M)$ becomes more complex as it has more non-zero weights. This increasing complexity is reflected by the fact that the loss function $M \mapsto L(w^M, \Xc_t)$ is a decreasing function of $M$. That is, the training error decreases as $M$ increases; see also the red curves in Figure \ref{fig:learning_curve_amorphous}. 

However, it is well known in machine learning and statistical learning theory that more complex models do not necessarily generalize better. The optimal regularization, here controlled by the sparsity parameter $M$, is determined via cross-validation using the validation set $\Xc_v$. That is, for each $1 \leq M \leq M_{\max}$ we compute the loss of the model $\widetilde{E}(x; w^M)$ on the validation set, and select the $M^{\star}$ that minimizes the validation error:
\begin{equation} \label{eqn: M cross validation}
    M^{\star} = \arginf_{1 \leq M \leq M_{\max}} L (w^M, \Xc_v) \, .
\end{equation}
The model that is used on the test data is $\widetilde{E}(x; w^{M^{\star}})$. In general $M^{\star} \neq M_{\max}$, since unlike the training error $M \mapsto L (w^M, \Xc_t)$, the validation error $M \mapsto L(w^M, \Xc_v)$ is not monotonically decreasing but rather generically decreases up to $M^{\star}$ and then increases after $M^{\star}$; see the green dashed curves in Figure \ref{fig:learning_curve_amorphous}. We remark that the value $M^{\star}$ is the best estimate of the optimal model for testing on states similar to those in the validation set; in other words, it balances the model between under-fitting and over-fitting. However, states for the extrapolation tasks (Section \ref{sec: data gen for extrapolation}) are not necessarily similar to the states in the validation set. Models that extrapolate must be formulated in such a way that when trained, the cross validation procedure selects a complexity $M^{\star}$ that captures underlying physical phenomena while ignoring non-physical patterns in the training and validation data. We achieve such a result by leveraging the universal wavelet scattering features (see Section \ref{subsec:ML_algorithm}), and by careful partitioning of the training and validation sets that is described in more detail in Section \ref{subsec:dataset_results}.

Computationally, solving \eqref{eqn: M-sparse weights} is NP-Hard. We thus solve a relaxed problem that obtains the weights in a greedy fashion, using the same orthogonal least squares (OLS) approach described in Hirn,  Mallat,  and  Poilvert~\cite{hirn:waveletScatQuantum2016}.  While the resulting weights do not in general solve \eqref{eqn: M-sparse weights} due to the relaxation, the OLS approach is optimal among greedy approaches since it reduces the mean square error by the maximum amount with each greedy step. Otherwise, using a greedy approach has two benefits. First, it is significantly more efficient; solving \eqref{eqn: M-sparse weights} requires $O(\binom{d}{M})$ floating point operations whereas the greedy approach requires $O(dM)$ floating point operations. Second, it is an iterative process that can be solved using a QR factorization, which in this case means that after $M$ nonzero weights are selected, the computation for $M+1$ nonzero weights requires solving only for the one additional weight. This lets us efficiently construct an array of models for $1 \leq M \leq M_{\max}$, which in turn enables an efficient solution to the cross validation problem given in \eqref{eqn: M cross validation}. 

Finally, we augment the learning process by leveraging empirical bootstrapping and feature bagging. Given an initial database of $n$ states and their energies (that does not include the withheld testing set), the empirical bootstrap algorithm samples the database with replacement to obtain the training set. Those states not selected for the training set are placed in the validation set. This approach allows us to construct many different models from one database, which are then averaged. The resulting averaged model, which is still a linear model over the representation $\Phi (x)$, is superior to any one individually fitted model since the averaging reduces random fluctuations in the fitting process that result from spurious patterns in a single training set. In order for this averaging process to have maximum effect, the weights of the individual models must be as uncorrelated as possible. Feature bagging, which is a prominent component of random forests, decorrelates the models by restricting the greedy selection at each greedy step. In particular, at each greedy step in the OLS algorithm, approximately $\sqrt{d + 1}$ features are sampled without replacement from among the full set of $d$ features in $\Phi (x)$ plus the bias term, minus the features that have already been selected up to that point. The OLS algorithm at each step must then select from among the sampled features, which due to the randomness in the feature sampling, results in models that are significantly less correlated. Indeed, in our own numerical experiments, the most significant features selected with empirical bootstrapping, but without feature bagging, are very often identical. While restricting the number of possible features at each greedy step means that each model has larger error on the training set, the aggregated average model improves on the test set~\cite{schwab:highBiasLowVarML2018}. 

\subsection{Atomic orbital wavelet scattering}
\label{subsec:ML_algorithm}

We now describe how we construct the feature vector $\Phi (x)$. Since our regression $\widetilde{E}(x; w) = w_0 + \langle w, \Phi (x) \rangle$ is a linear model over $\Phi (x)$, the representation $\Phi (x)$ should have the same properties as $E (x)$. In particular, as has been noted by several machine learning papers for many-particle physics~\cite{bartok:repChemEnviron2013}, $E(x)$ is invariant to translations, rotations, and reflections (i.e., isometries) of the atomic coordinates $\{ R_k \}_{k=1}^{N_x}$, and therefore $\Phi (x)$ should also be invariant to isometries. Additionally, $E(x)$ is independent of the atom indexation in the list $x$, and thus $\Phi (x)$ must be invariant to index permutations. Like the formation energy, $\Phi (x)$ should be a continuous function of the atomic coordinates $\{ R_k \}_{k=1}^{N_x}$, which is particularly important since our data consists of structural relaxation paths. Furthermore, since we are fitting total energy per atom, the feature values should be independent of the number of atoms in the unit cell. Finally, the amorphous $\text{Li}_{\alpha} \text{Si}$ systems are periodic, and thus the features $\Phi (x)$ must be invariant to equivalent representations of $x$ with respect to the periodicity of the state.

In addition to those basic physical properties, electronic interactions encoded by the molecular Hamiltonian are inherently multiscale in nature and thus range over a number of length scales. The resulting total energy of the system is a nonlinear function of these length scales. We thus seek a representation $\Phi (x)$ that on the one hand can separate the length scales of the system, while on the other hand can recombine them in a nonlinear fashion. Our approach is to use the atomic orbital wavelet scattering transform of Brumwell \emph{et al}.,\cite{brumwell2018steerable} which itself is an adaptation of the three-dimensional solid harmonic wavelet scattering transform proposed by 
~\citet{eickenberg:3DSolidHarmonicScat2017, eickenberg:scatMoleculesJCP2018} for molecules. We review its construction in this section, emphasizing certain nuances specific to atomic states with periodic unit cell and more specifically to $\text{Li}_{\alpha} \text{Si}$ systems.

The wavelet scattering transform is based upon the wavelet transform of three-dimensional signals. We identify the state $x$ with such a signal, which will encode permutation invariance into all representations derived from this signal. Let $\Qc_x \subset \R^3$ be the unit cell of the state $x$, which in the case of all systems we consider is a cube. We encode the state $x$ as a superposition of Dirac delta functions:
\begin{equation*}
    \rho_x^f (u) = \sum_{k=1}^{N_x} f(Z_k) \delta (u - R_k) \, , \quad u \in \Qc_x \, .
\end{equation*}
We use the notation $\rho_x^f$ because one can think of it as a type of nuclear density for the state $x$, but in which we allow some additional flexibility. In particular, the function $f: \N \rightarrow \R$ encodes a weight that responds based on the type of atom. We use five different functions $f$, which can be thought of as channels of the state $x$, similar to how a color image has red, green, and blue channels. The five channels we use are lithium, silicon, valence, ionic, and kinetic. The lithium and silicon channels partition the state $x$ along atom species, whereas the valence and core channels separate the state $x$ according to electron type. The kinetic channel, inspired by the Thomas--Fermi--Dirac--von Weizsacker model in quantum chemistry, encodes a different scaling in the number of electrons than the other four channels. The precise definitions of these channels are given in Appendix \ref{sec: channel def}.

A simple translation and rotation invariant representation of $x$ is obtained by summing over $f_x = (f(Z_k))_{k=1}^{N_x}$ for each channel $f$: 
\begin{equation} \label{eqn: zero order features}
\phi_{\gamma} (x) = N_x^{-1} \| f_x \|_q^q = \frac{1}{N_x} \sum_{k=1}^{N_x} |f (Z_k)|^q \, , \quad \gamma = (f, q) \, .
\end{equation}
We compute four different types of summations by taking $q^{\text{th}}$ powers of $f(Z_k)$ for 
\begin{equation} \label{eqn: q values}
    q \in \{1, 4/3, 5/3, 2 \} \, .
\end{equation} 
These powers are also inspired by the Thomas--Fermi--Dirac--von Weizsacker model, and will be discussed more later in this section. By dividing by $N_x$, the features $\phi_{\gamma} (x)$ are also invariant to system size. Finally, since they are constant with respect to the atomic coordinates, they are trivially continuous functions of them.

\begin{figure}
    \centering
    \includegraphics[trim=0.35in 0.35in 0.35in 0.35in, clip,width=0.25\linewidth]{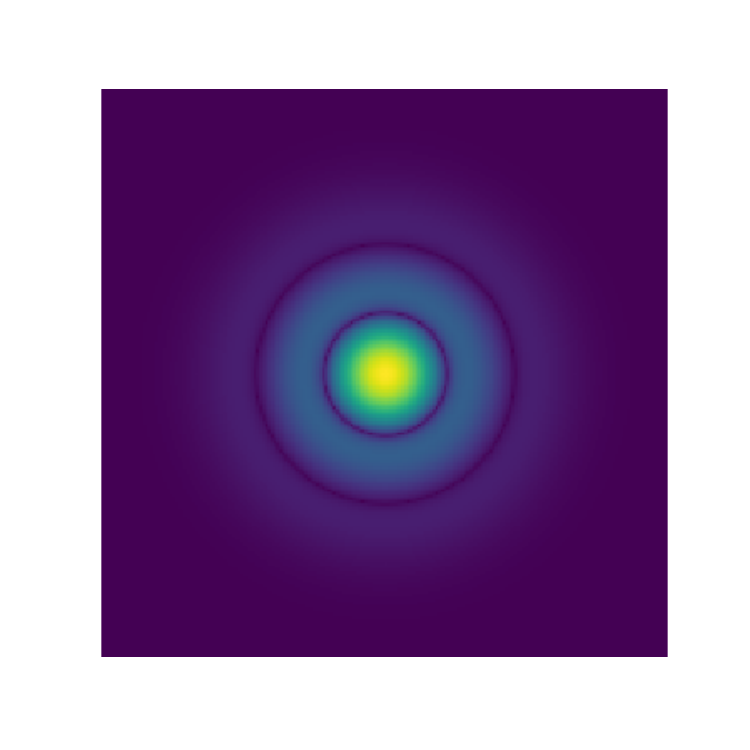}
    \includegraphics[trim=0.35in 0.35in 0.35in 0.35in, clip,width=0.25\linewidth]{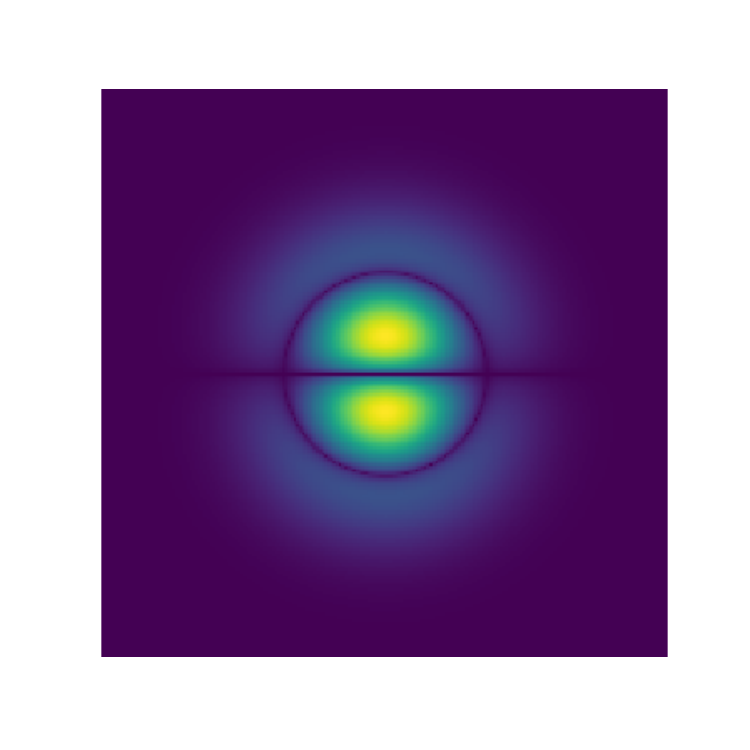}
    \includegraphics[trim=0.35in 0.35in 0.35in 0.35in, clip,width=0.25\linewidth]{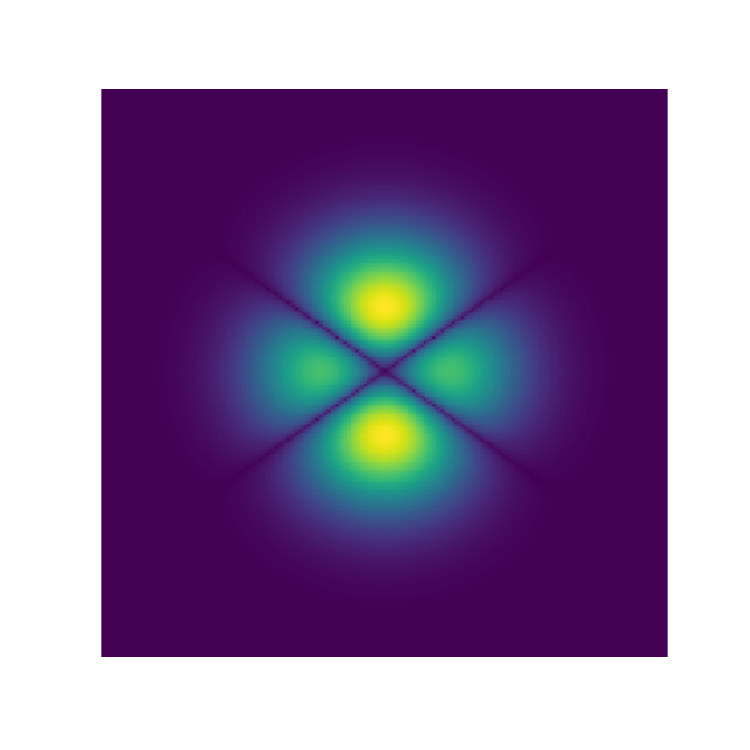}\\
    \includegraphics[trim=0.35in 0.35in 0.35in 0.35in, clip,width=0.25\linewidth]{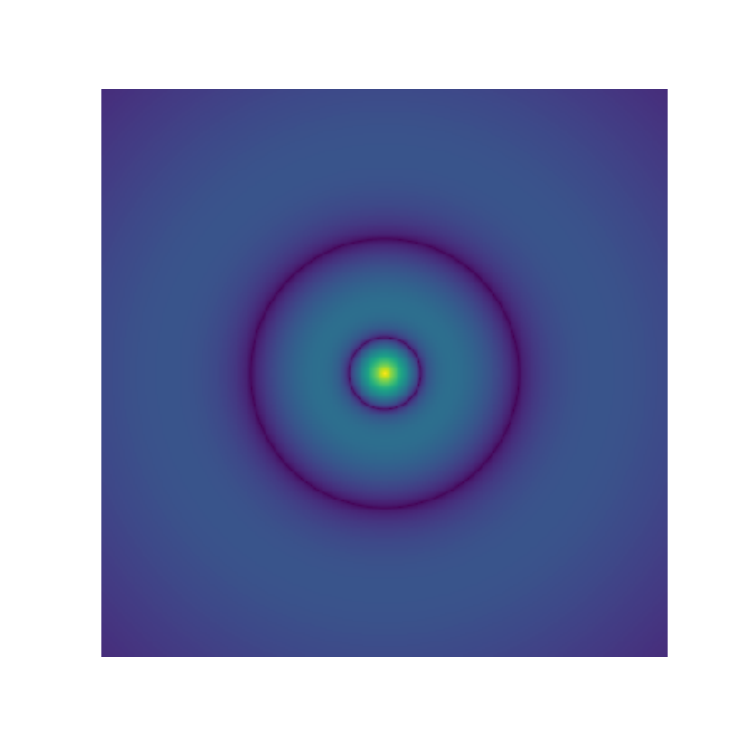}
    \includegraphics[trim=0.35in 0.35in 0.35in 0.35in, clip,width=0.25\linewidth]{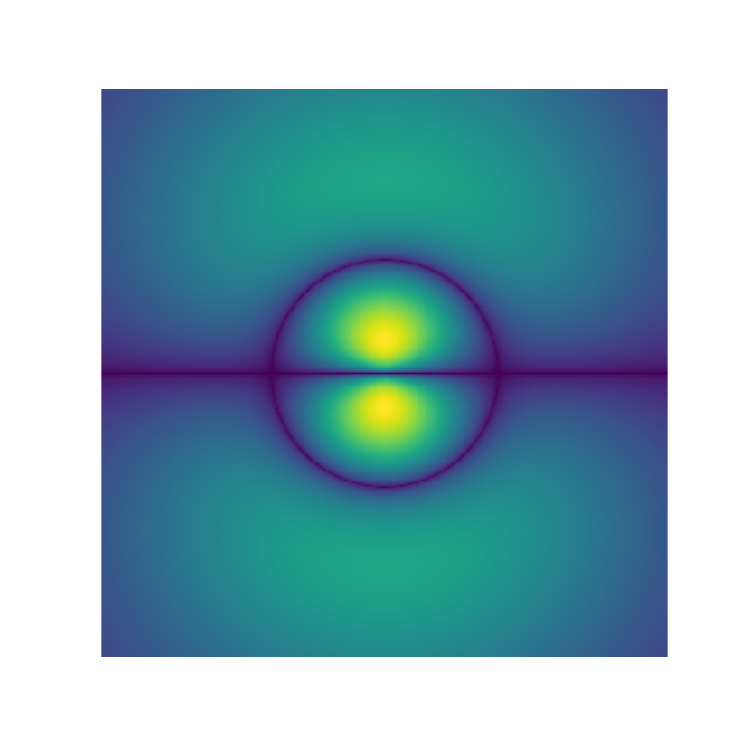}
    \includegraphics[trim=0.35in 0.35in 0.35in 0.35in, clip,width=0.25\linewidth]{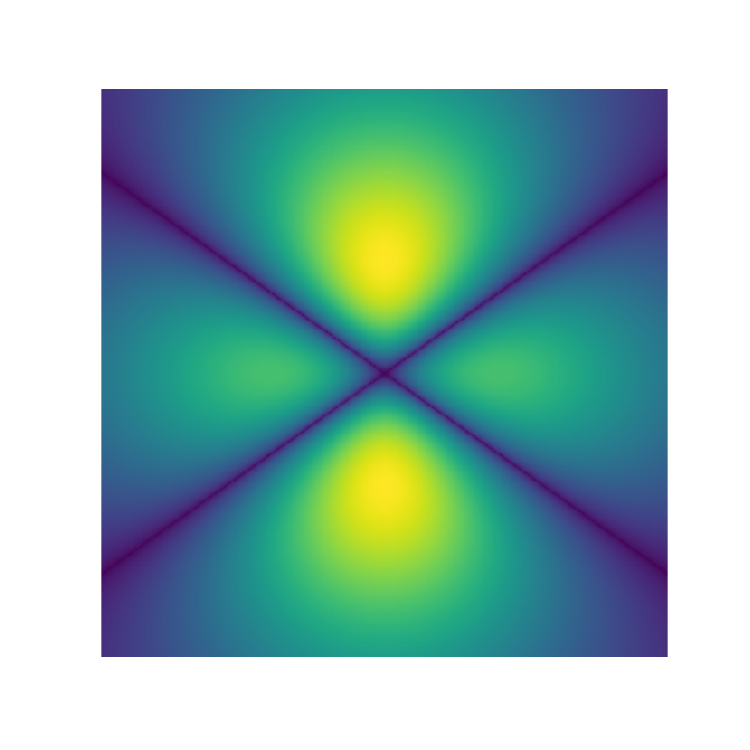}
    \caption{Top row left to right: Density plot cross sections in the $xz$-plane of atomic orbital wavelets for $(n, \ell, m)=(3,0,0)$, $(3,1,0)$, and $(3,2,0)$.  Bottom Row: Corresponding plots for the hydrogen atom orbitals  ($3s$, $3p$, and $3d$ orbitals, respectively). All images are rescaled for visualization. Note that the exponential radial decay $e^{-r}$ of the hydrogen orbitals is replaced by a Gaussian decay $e^{-r^2}$ in the wavelets accounting for the greater localization of the wavelets.}
   \label{fig:wavelets_and_orbitals}
\end{figure}

The zero order features \eqref{eqn: zero order features} satisfy all the required invariance properties, but they remove all geometric information from the state $x$ and are constant for a given lithium-silicon ratio $\alpha$. We compute complimentary features that separate the length scales of $x$ and encode multiscale geometric invariants. These features will be derived from a three-dimensional wavelet transform of $\rho_x^f$, which gives a multiscale representation of the signal. Following Brumwell \emph{et al.}\cite{brumwell2018steerable}, we define a family of atomic orbital wavelets $\psi_{n,\ell}^m : \R^3 \rightarrow \C$, 
\begin{align*}
    \psi_{n,\ell}^m (u) = Q_{n,\ell} (|u|) &Y_{\ell}^m (u / |u|) \, , \\
    &n \geq 1 \, , \, 0 \leq \ell < n \, , \, |m| \leq \ell \, ,
\end{align*}
where $Y_{\ell}^m$ is the usual spherical harmonic function and $Q_{n,\ell}$ is a radial function defined as:
\begin{equation} \label{eqn: radial fcn}
    Q_{n, \ell}(r) = C_{n, \ell} r^{\ell} L_{n - \ell - 1}^{\ell + 1/2} \left( \frac{r^2}{2\beta^2} \right) e^{-r^2 / 2\beta^2} \, , \quad r \geq 0 \, .
\end{equation}
Here $C_{n, \ell}$ is a normalizing constant and the functions $L_k^{\nu}$ are the associated Laguerre polynomials. We refer to the family of wavelets $\psi_{n, \ell}^m$ as atomic orbital wavelets since they mimic the shape of the hydrogen atomic orbitals. Indeed, $(n, \ell) = (1,0)$, $(2,0)$, $(2,1)$ corresponds to the 1s, 2s, and 2p orbitals, respectively, with similar correspondences for larger values of $n$; see Figure \ref{fig:wavelets_and_orbitals}. While the hydrogen atomic orbitals have exponential scaling, here we use a Gaussian function, which mimics the well-known Gaussian type orbitals from the quantum chemistry literature. 

We use these wave functions as wavelets, though, in which the wavelet transform dilates the wavelet at different dyadic scales $2^j$,
\begin{equation*}
    \psi_{j, n, \ell}^m (u) = 2^{-3j} \psi_{n, \ell}^m (2^{-j} u) \, , \quad 0 \leq j < J \, ,
\end{equation*}
which increases the size of the wavelet. Let $s_x$ be the side length of the cubic unit cell $\Qc_x$. Unlike the molecular systems studied by ~\citet{eickenberg:3DSolidHarmonicScat2017, eickenberg:scatMoleculesJCP2018} here $x$ is a periodic system and so we compute a periodic wavelet transform of the density $\rho_x^f$ using a circular convolution $\circledast$ defined as:
\begin{equation} \label{eqn: circular convolution}
    \rho_x^f \circledast \psi_{j, n, \ell}^m (u) = \sum_{p \in \Z} \rho_x^f \ast \psi_{j, n, \ell}^m (u - p s_x) \, .
\end{equation}
The operation $\ast$ is the usual convolution over $\R^3$, which for the nuclear density-type function $\rho_x^f$ yields:
\begin{equation} \label{eqn: regular convolution}
    \rho_x^f \ast \psi_{j, n, \ell}^m (u) = \sum_{k=1}^{N_x} f(Z_k) \psi_{j, n, \ell}^m (u - R_k) \, .
\end{equation}

Examining \eqref{eqn: circular convolution} and \eqref{eqn: regular convolution} we have the following interpretations. The standard convolution $\rho_x^f \ast \psi_{j, n, \ell}^m$ emits the wavelet $\psi_{j, n, \ell}^m$ from the location of each atom in the unit cell of $x$, with a strength given by $f(Z_k)$. The interference patterns encoded by these emissions encode geometric information of the state of the system at different scales $2^j$, which we shall aggregate to form multiscale, invariant features. The circular convolution $\rho_x^f \circledast \psi_{j, n, \ell}^m$ wraps the wavelets periodically in the unit cell $\Qc_x$, thus giving us a periodic function that respects the periodicity of the system. The parameter $\beta$ in \eqref{eqn: radial fcn} in the definition of $Q_{n,\ell}$, which encodes the smallest wavelet scale, is selected so that $\psi_{n, \ell}^m (u - R_k)$ and $\psi_{n, \ell}^m (u - R_l)$ interfere only if $|R_k - R_l|$ is small, i.e., if the atoms located at $R_k$ and $R_l$ are neighboring atoms. The maximum scale $2^{J-1}$ is selected so that the size of the wavelet $\psi_{J-1, n, \ell}^m$ is on the order of the maximum side length $s_x$ of the unit cells $\Qc_x$ across all training states, thus enabling the corresponding wavelet coefficients $\rho_x^f \circledast \psi_{J-1, n, \ell}^m$ to encode macroscopic patterns in the arrangement of the atoms in $x$. These choices allow us to capture short-range interactions in features derived from wavelet filters with small $j$, while wavelet filters with large $j$ capture interactions across a larger span of the system.

\begin{figure}[tp]
    \centering
    \includegraphics[trim=1in 0.4in 0.7in 0.4in, clip,width=0.9\columnwidth]{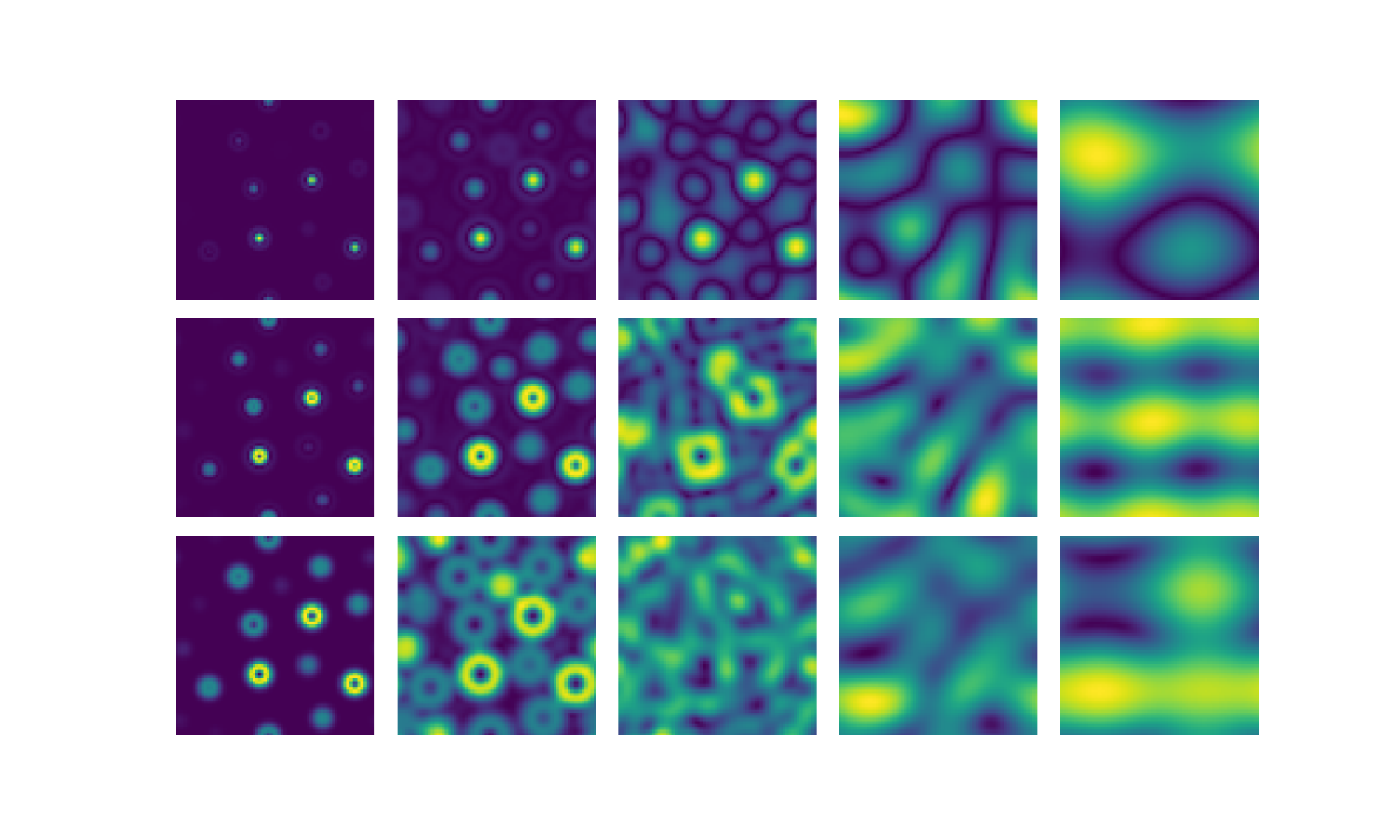}
    \caption{Cross-sections of the first order nonlinear, equivariant maps $\sigma (\rho_x^f \circledast \psi_{j, n, \ell}) (u)$. The $\log_2$ scales  $j = 0,1,2,3,4$ increase from left to right, respectively, and the angular quantum number $\ell = 0,1,2$ from top to bottom, respectively, with $n=3$. The maps extract multiscale geometric information on the arrangement of the atoms in the state $x$.}
    \label{fig:first order nonlinearity}
\end{figure}

Convolution operators are, in general, translation equivariant but not rotation equivariant. However, the atomic orbital wavelet filters are designed to admit a rotationally equivariant representation by combining information across the magnetic quantum number $m$ through the following nonlinear transform $\sigma$:
\begin{equation*}
    \sigma (\rho_x^f \circledast \psi_{j, n, \ell}) (u) = \left( \sum_{m = -\ell}^{\ell} \abs*{\rho_x^f \circledast \psi_{j, n, \ell}^m(u)}^2 \right)^{1/2}.
\end{equation*}
The collection of maps $\sigma (\rho_x^f \circledast \psi_{j, n, \ell})$ constitutes a multiscale, isometry equivariant representation of the state $x$; see Figure \ref{fig:first order nonlinearity} for plots of these maps. We note that the wavelet filters with $\ell = 0$ are radially symmetric and therefore their convolutions are equivariant to rotations with or without the nonlinear transform $\sigma$.

Equivariant representations yield invariant representations via integral operators that integrate over the space variable $u$. We compute $\Lb^q (\Qc_x)$ norms, for the same four $q$ values in \eqref{eqn: q values}, of the maps $\sigma (\rho_x^f \circledast \psi_{j, n, \ell})$:
\begin{equation} \label{eqn: norms}
    \| \sigma ( \rho_x^f \circledast \psi_{j, n, \ell} ) \|_q^q = \int_{\Qc_x} | \sigma (\rho_x^f \circledast \psi_{j, n, \ell}) (u)|^q \, du \, .
\end{equation}
The selection of powers $q$ is motivated by the Thomas--Fermi--Dirac--von Weizsacker model in quantum chemistry, in which the $4/3$ scaling is used to approximate the exchange energy, the $5/3$ scaling is used to approximate the kinetic energy, and the power of $2$ encodes an additional part of the kinetic energy and pairwise Coulombic interactions (see also ~\citet{hirn:waveletScatQuantum2016}). The power $q=1$ is also used since these integrals scale linearly with $\sum_k f(Z_k)$. 

We normalize the norms \eqref{eqn: norms} to be invariant to system size, which defines first order wavelet scattering features:
\begin{equation*}
    \phi_{\gamma} (x) = N_x^{-1} \| \sigma ( \rho_x^f \circledast \psi_{j, n, \ell} ) \|_q^q \, , \quad \gamma = (f, q, j, n, \ell) \, .
\end{equation*}
In numerical experiments reported on in Section \ref{sec:results}, there are five channels $f$, five scales $j$ (i.e., $J = 5$), $n = 3$ (that is, a single $n$ is used), $0 \leq \ell < n = 3$, and there are the four $q$ values specified in \eqref{eqn: q values}, which yields 300 first order features. The smallest wavelet scale (at $j=0$) is set with the parameter $\beta$ in \eqref{eqn: radial fcn} to be large enough to ensure nontrivial overlap between nearest neighbor atoms in the database, which corresponds to a length scale of $0.9$ \AA. The largest length scale (at $j = J-1 = 4$) is chosen so that the wavelet supports envelop the largest unit cell in the database which has side length $11.9$ \AA.  These first order wavelet scattering features encode isometry and size invariant descriptions of the state $x$ across multiple length scales $2^j$ for $0 \leq j < J$. Furthermore, since the atomic orbital wavelets $\psi_{n, \ell}^m$ are continuous functions, the resulting maps $\sigma (\rho_x^f \circledast \psi_{j, n, \ell})$ are continuous functions of the atomic coordinates, which means their integrals are as well. The use of circular convolution ensures the maps are invariant to the representation of $x$ with respect to its periodicity. 

\begin{figure}[tp]
    \centering
    \includegraphics[trim=1in 0.4in 0.7in 0.4in, clip,width=0.9\columnwidth]{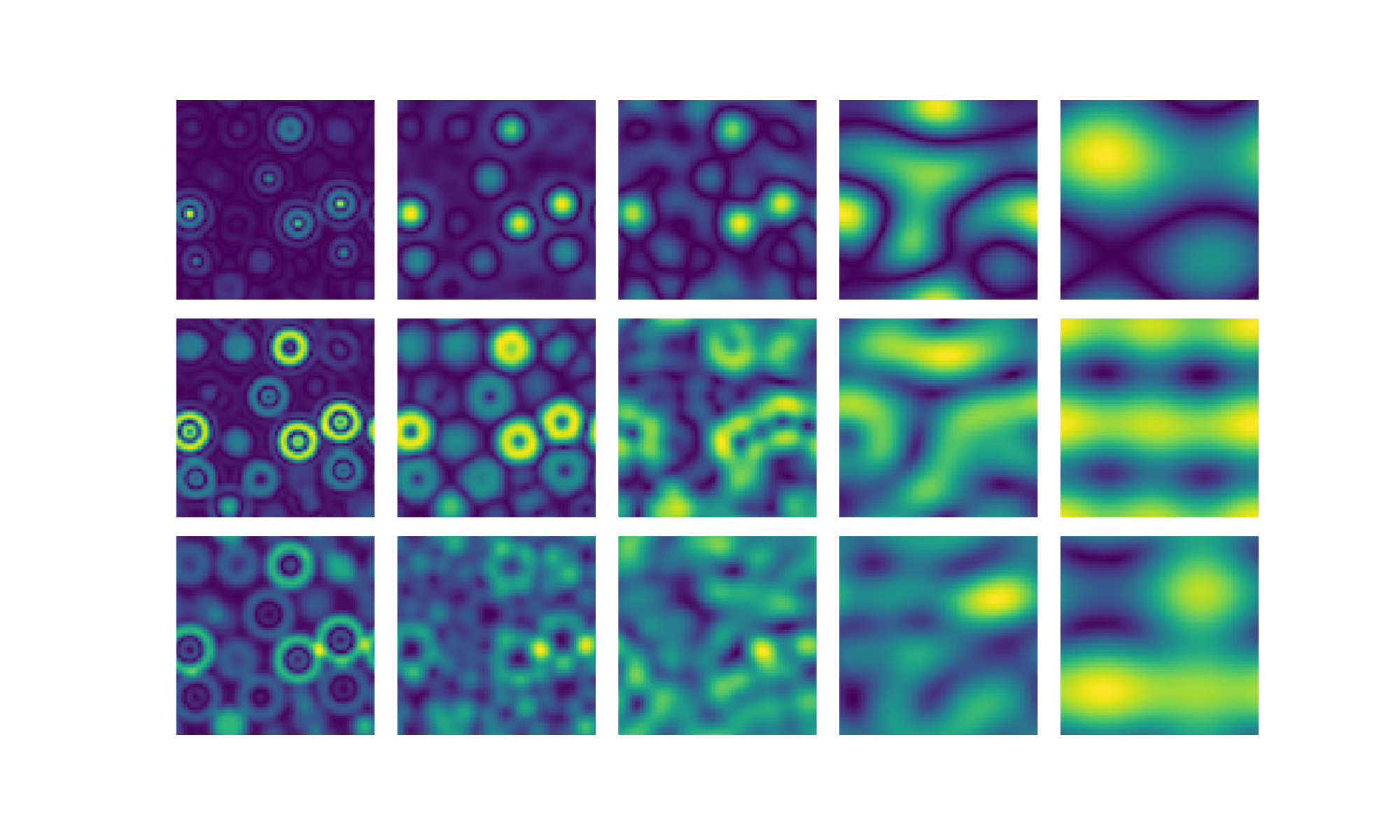}
    \caption{Cross-sections of the second order nonlinear, equivariant maps $\sigma(\sigma(\rho_x^f \circledast \psi_{j_1, n_1, \ell_1}) \circledast \psi_{j_2, n_2, \ell_2}) (u)$ for $(j_1,n_1,\ell_1)=(1,3,1)$, which is the second from the top and second from the left in Figure \ref{fig:first order nonlinearity}. The $\log_2$ scales $j_2$ and $\ell_2$ vary the same as in Figure \ref{fig:first order nonlinearity}, and $n_2=3$. Notice that many of the multiscale geometric patterns are distinct from those in Figure \ref{fig:first order nonlinearity}.}
    \label{fig:second order nonlinearity}
\end{figure}

First order wavelet scattering features are complemented by second order wavelet scattering features that incorporate multiple length scales of $x$ into a single feature. They are computed by iterating the nonlinear wavelet transform, which couples the scales $2^{j_1}$ and $2^{j_2}$:
\begin{align}
    &\sigma(\sigma(\rho_x^f \circledast \psi_{j_1, n_1, \ell_1}) \circledast \psi_{j_2, n_2, \ell_2}) (u) = \nonumber \\
    &\left(\sum_{m = -\ell_2}^{\ell_2} \abs*{\sigma (\rho_x^f \circledast \psi_{j_1, n_1, \ell_1}) \circledast \psi_{j_2, n_2, \ell_2}^{m}(u)}^2 \right)^{1/2} \, . \label{eqn: second order equivariant maps}
\end{align}
The second order maps \eqref{eqn: second order equivariant maps}, which resemble the architecture of a convolutional neural network as well as aspects of tensor field networks~\cite{smidt:tensorFieldNetworks2018}, are equivariant with respect to translations and rotations, and extract coupled geometric information at the scales $2^{j_1}$ and $2^{j_2}$ from the state $x$. Figure \ref{fig:second order nonlinearity} plots examples of these maps, which are noticeably different than their first order counterparts.

Second order invariant wavelet scattering features are computed analogously to the first order features, by taking normalized $\Lb^q (\Qc_x)$ norms of the equivariant maps:
\begin{align}
    \phi_{\gamma} (x) &= N_x^{-1} \norm*{\sigma \left(\sigma (\rho_x^f \circledast \psi_{j_1, n_1, \ell_1}) \circledast \psi_{j_2, n_2, \ell_2} \right)}_q^q \nonumber \\ 
    &= \frac{1}{N_x} \int_{\Qc_x} \abs*{\sigma \left(\sigma( \rho_x^f \circledast \psi_{j_1, n_1, \ell_1}) \circledast \psi_{j_2, n_2, \ell_2} \right)(u)}^q \, du \nonumber \\
    \gamma &= (f, q, j_1, n_1, \ell_1, j_2, n_2, \ell_2) \, . \nonumber 
\end{align}
Using the same parameters as the first order features, plus setting $n_2 = 3$ and $0 \leq \ell_2 < n_2 = 3$ and $\max(0, j_1 - 1) \leq j_2 < J = 5$, we see there are 3420 second order wavelet scattering features. They thus greatly expand the representation of the state $x$, and satisfy all the required invariance properties. 

We collect the zero, first, and second order invariant wavelet scattering features into a single feature representation $\Phi (x) \in \R^{3740}$. Let $\lambda = (j,n,\ell)$ denote the triplet of wavelet parameters. With this feature representation, our energy model \eqref{eqn: ml energy model} can be written as:
\begin{align}
    \widetilde{E}_{\text{tot}}^{\ast} &(x; w) = \widetilde{E} (x; w) = \nonumber \\
    ={} &w_0 + \frac{1}{N_x} \sum_{f, q} w_{f,q} \| f_x \|_q^q \nonumber \\
    &{}+ \frac{1}{N_x} \sum_{f, q, \lambda} w_{f, q, \lambda} \| \sigma (\rho_x^f \circledast \psi_{\lambda}) \|_q^q \label{eqn: scattering energy model} \\
    &{}+ \frac{1}{N_x} \sum_{f, q, \lambda_1, \lambda_2} w_{f, q, \lambda_1, \lambda_2} \| \sigma ( \sigma (\rho_x^f \circledast \psi_{\lambda_1}) \circledast \psi_{\lambda_2}) \|_q^q \, , \nonumber 
\end{align}
with the weights solved for using the algorithms described in Section \ref{sec: linear regression and model fitting}. We remark that both the features themselves, and the number of features, constitute a vast simplification over the model presented in Brumwell \emph{et al}~\cite{brumwell2018steerable}.

\section{Numerical Results}
\label{sec:results}

We report empirical errors for formation energy prediction and related tasks. Section \ref{subsec:dataset_results} presents interpolation type test errors on predicting relaxation path formation energies for amorphous $\text{Li}_{\alpha} \text{Si}$ of a similar nature to those in the training set. Sections \ref{subsec:diffusion_results}, \ref{subsec:large_results}, and \ref{subsec:modulus_results} report extrapolation task errors, focusing on diffusion barrier estimation, formation energies of amorphous states with larger unit cells, and bulk modulus prediction, respectively. Appendix \ref{sec: alternative models} compares these results with results obtained by varying the training procedures and model formulation presented in Section \ref{sec:methods}. In particular, it considers a non-randomized training procedure that does not utilize bootstrapping and feature bagging (originally described in Section \ref{sec: linear regression and model fitting}), as well as an energy model that utilizes only zero and first order wavelet scattering features (defined in Section \ref{subsec:ML_algorithm}). In each case, we see the advantage of the full algorithm. 

\begin{table}[t]
\begin{tabular}{|l|l|l|}
\hline \hline
& RMSE (meV/atom) & MAE (meV/atom) \\ \hline
Relaxation paths & 7.44 $\pm$ 0.49 & 5.52 $\pm$ 0.34 \\ 
Diffusion & 12.3 $\pm$ 0.50 & 11.7 $\pm$ 0.51 \\ 
Large states & 9.54 $\pm$ 0.25 & 6.81 $\pm$ 0.23 \\ 
Bulk modulus & 12.8 $\pm$ 1.36 & 8.92 $\pm$ 0.68 \\ \hline
\hline
\end{tabular}
\caption{Numerical results for ML predictions on the test data from the amorphous dataset and the three extrapolation tasks from the model trained only on the amorphous data.}
\label{table: energy prediction errors}
\end{table}

\subsection{Training and testing on amorphous dataset}
\label{subsec:dataset_results}

Recall from Section \ref{subsec:training_data_generation} and Figure \ref{fig:dataset} we have a training database of 90,252 amorphous $\text{Li}_{\alpha} \text{Si}$ structures with DFT computed energies spread across 37 concentrations $\alpha$, ranging from 0.1 to 3.75. These 90,252 structures correspond to 370 relaxation paths of an initial set of 370 high energy states, with 10 relaxation paths per concentration. Unlike many standard machine learning approaches which obtain a training set by uniformly sampling data points from the initial database, here we uniformly randomly sample relaxation paths. Using five-fold cross validation, we randomly partition the relaxation paths into five sets of 74 relaxation paths with two paths per concentration in each of the sets of 74. We place four of these sets, 296 relaxation paths total, in the training/validation set, and one set of 74 paths in the test set. We rotate through using each set separately as a test set, meaning that we carry out all numerical experiments five times, each time with a different training/validation and test set split.

We select the training/validation/testing sets according to relaxation data paths, and not structures, because it leads to more physically realistic training and testing scenarios. In particular, it is reasonable to assume that whole relaxation paths, computed via DFT, would be used to train a model which is then used to compute relaxation paths of new high energy states. Here the validation and testing sets are simpler in that we require the model to predict all formation energies along a new relaxation path in which the structures along the path are given. Nevertheless, empirical results indicate this training paradigm significantly restricts the degree to which the machine learned model can fit non-physical spurious patterns in the data. We leave for future work developing a model that can predict the entire relaxation path starting with the only the highest energy state. 

\begin{figure}[t]
    \centering
    \includegraphics[width=0.7\columnwidth]{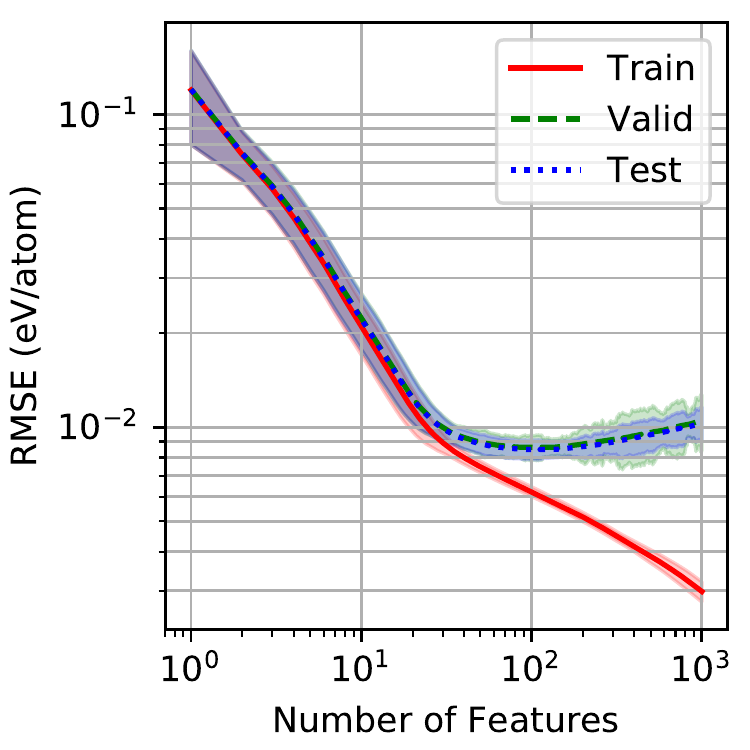}%
    \caption{Errors on the amorphous \ce{Li_{\alpha}Si} database as a function of number of features included in the model on a log-log scale.  Error on the training set is shown in red, the validation set is shown in green, and the test set is shown in blue. The training error is a decreasing function of the number of features, whereas the validation and testing curves are not. The value $M^{\star}$ that minimizes the validation curve is the algorithm's best estimate for the optimal model that best balances under- and over-fitting of the training data. It has good agreement with the minimum of the test error curve.}
    \label{fig:learning_curve_amorphous}
\end{figure}

Using the 296 training relaxation paths, we carry out the model fitting algorithm described in Section \ref{sec: linear regression and model fitting}. For the training set, we randomly select according to a uniform distribution, with replacement, 296 relaxation paths from the training set. Those paths selected more than once are repeated in the training set with the number of copies equalling the number of times the path was selected. Those paths that are not selected are placed in the validation set. The sparse linear model is trained using the greedy OLS algorithm with randomized feature bagging, with the number of features $M$ ranging from $M = 1$ to $M = M_{\max} = 1000$. The optimal number of features $M = M^{\star}$ is selected by minimizing the loss on the validation set. This procedure is repeated 100 times, resulting in 100 sparse linear models of the form \eqref{eqn: scattering energy model}, which are averaged together to yield the final model. 

\begin{figure}[tp]
\includegraphics[width=0.7\columnwidth]{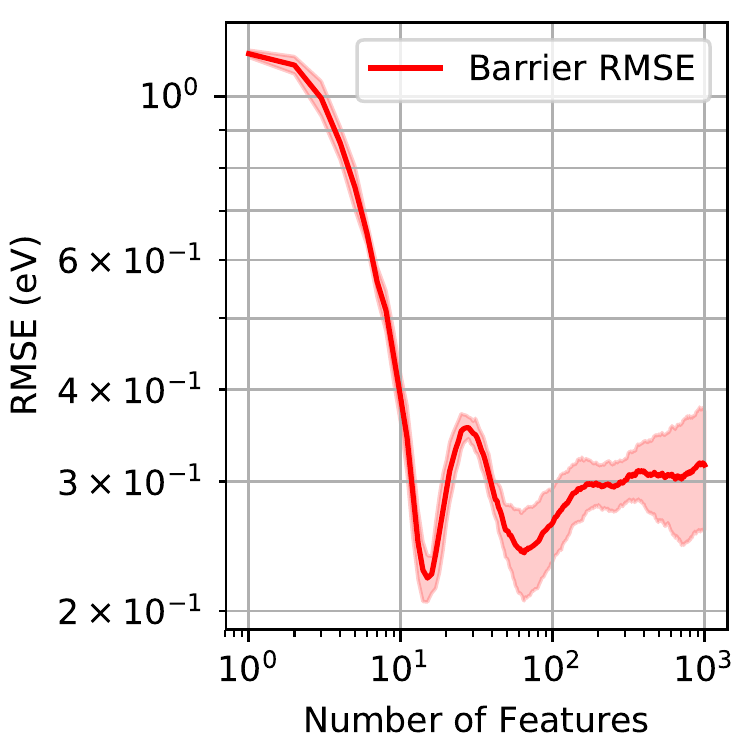}%
\caption{\label{fig:TranStatesRMSE} Log-log plot of RMSE in diffusion barrier prediction averaged over the five folds.}%
\end{figure}

This final model is evaluated on the withheld test set. Figure \ref{fig:learning_curve_amorphous} depicts the training, validation, and testing errors as a function of the number of model features $M$. It indicates that best models have, generally, between 64--256 features, with an average of 121 features per model, a small number given that there are approximately 70,000 training structures. Furthermore, the validation curve closely follows the test curve, indicating that our cross-validation procedure is nearly optimal for this test data. The average root mean squared error (RMSE) and the average mean absolute error (MAE) over the five test folds, along with the standard deviation, is reported in the first row (relaxation paths) of Table \ref{table: energy prediction errors}. Despite the small number of features, the RMSE is 7.44 meV/atom and the MAE is 5.52 meV/atom, which is comparable to the results reported in ~\citet{onat:implantNNLiSi2018} and Artrith, Urban, and Ceder~\cite{Artrith:LiSiML2018}, both of which used neural networks, and is small enough to be of use in materials science applications. However, the model developed here is significantly simpler than neural network models, being a linear model over multiscale, invariant features that utilize a universal set of filters. As such, the model is adept at generalization, as reported in the next three subsections. 

\subsection{Extrapolate: Diffusion in amorphous system}
\label{subsec:diffusion_results}

One important application of atomistic simulation is the study of atomic migration from site to site.  The energetic barrier to migration determines diffusion constants and ionic conductivity~\cite{Sholl09}.  The diffusion process may be simulated by directly tracking the mean square displacement using molecular dynamics, or by calculating the migration barrier and using the Nernst-Einstein relationship. The first step in the explicit calculation is to find the minimum-energy path (MEP) for an atom to travel between two stable sites.  This is typically done using optimization techniques such as the Nudged Elastic Band (NEB) method~\cite{Johnsson98}.  The barrier is defined as the energy difference between the stable position and the highest-energy position (saddle point) along the MEP.

\begin{table}
    \centering
    \begin{tabular}{|c|c|c|}
    \hline \hline
    Path & Barrier (ML Model) & Barrier (DFT)\\  \hline 
1 & 0.228 & 0.226 \\
2 & 0.819 & 0.341 \\
3 & 2.256 & 2.139 \\
4 & 0.230 & 0.402 \\
5 & 2.613 & 2.224 \\
6 & 0.326 & 0.354 \\ \hline \hline 
    \end{tabular}
    \caption{Diffusion barriers (in eV) along various paths as predicted by our ML model and DFT.  Paths 1-5 start from the same Li\textsubscript{0.2}Si structure and path 6 is in Li\textsubscript{0.5}Si. }
    \label{tab:barriers}
\end{table}

There are a number of reasons why calculation of diffusion barriers may present a challenge for our ML model.  Our present models do not predict forces, so they cannot be used with NEB for prediction of the path itself.  We therefore simply predict energies along the DFT-calculated MEP.   A more fundamental challenge is the fact that the transition state structure, with one atom in a high-energy state and the rest in relatively low-energy states, is qualitatively different from the training items in the amorphous \ce{Li_{\alpha}Si} data set.  Calculation of diffusion barrier is thus an extrapolation task.  Furthermore, there is only one diffusing atom in the simulation box during calculation of the diffusion barrier.  This means that energy per atom is no longer the most relevant measure of error.  Instead, total energy differences between similar structures along the MEP are the relevant quantity.  Cancellation of systematic errors in DFT allows the calculation of energy differences along diffusion paths with much higher accuracy than would be suggested based on the accuracy of the method in total energy per atom~\cite{Sholl09}.  It remains to be seen if similar cancellation of errors can improve the accuracy of diffusion barriers predicted by an ML model.

\begin{figure*}[t]
\includegraphics[width=0.8\textwidth]{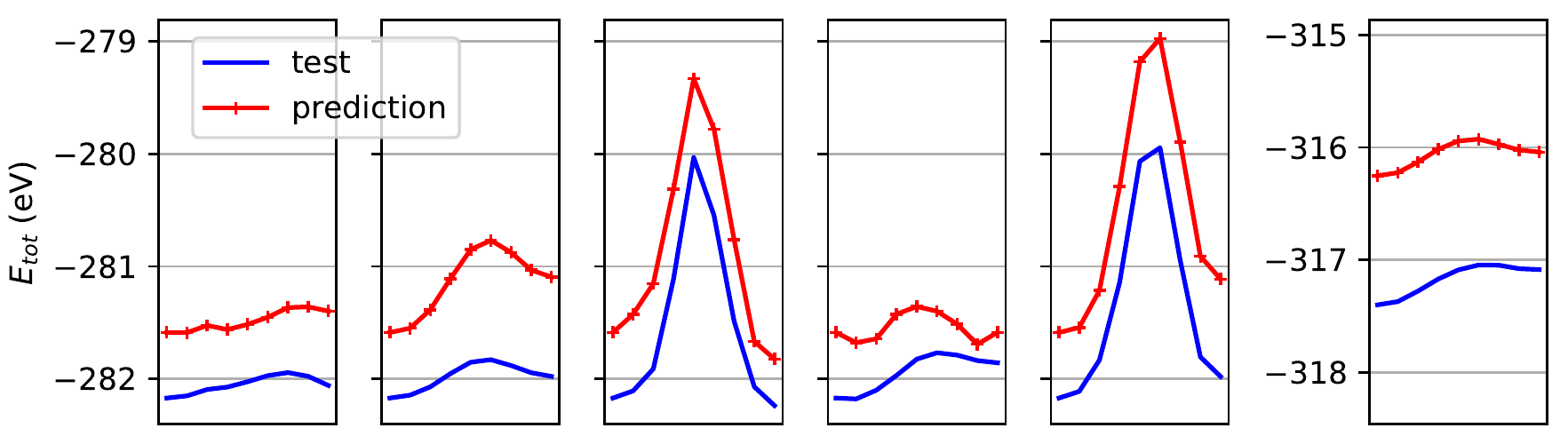}\\
\includegraphics[width=0.8\textwidth]{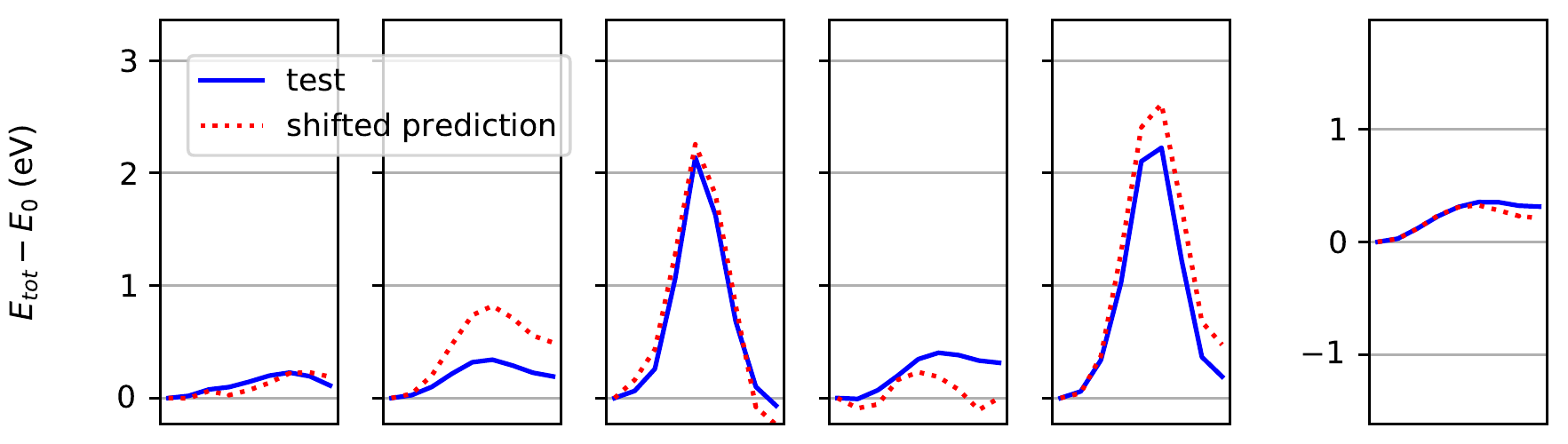}\\
\includegraphics[width=0.8\textwidth]{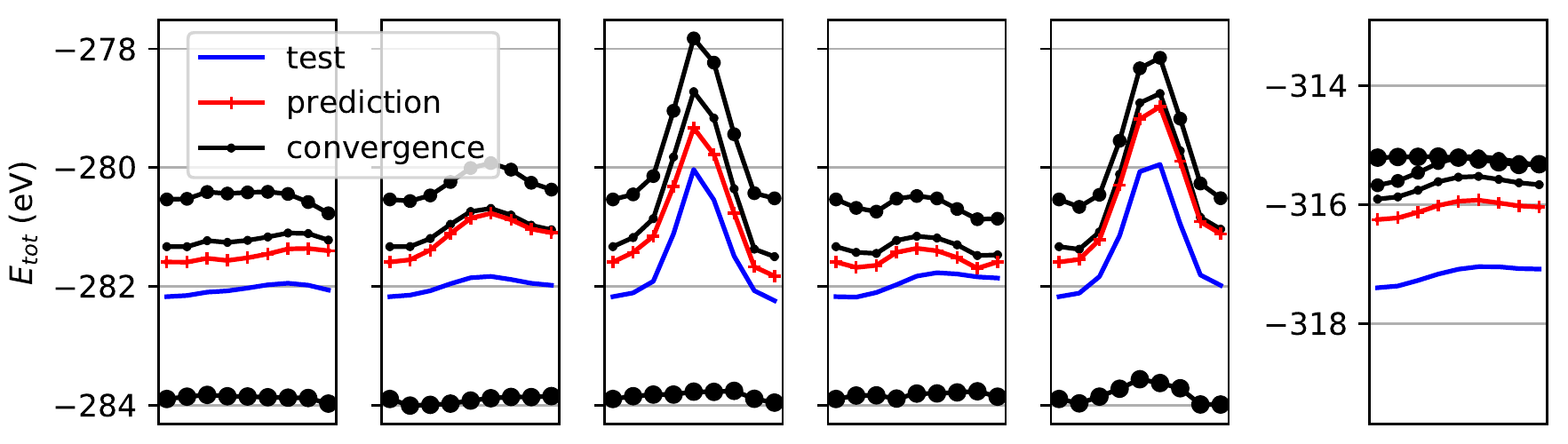}%
\caption{\label{fig:TranStates} Plots of the six diffusion barrier paths (blue) and (top row) model predictions in red, (middle row) model predictions and test data shifted by their respective starting-point energies $E_0$, and (bottom row) convergence of models with increasing number of features used for predictions of diffusion barrier curves. The large radii circles coincide with fewer features used starting from a model with a single feature. The models quickly converge in shape and progress towards the red curve which is the aggregate model prediction. There is a curve for each choice of number of features $M \in \{1, 21, 41\}$.  }%
\end{figure*}

To test the extrapolation of our model to diffusion barriers, void spaces were identified in \ce{Li_{0.2}Si} and \ce{Li_{0.5}Si} by Voronoi analysis.  Candidate endpoint structures were created by moving nearby lithium atoms into the voids and relaxing the resulting structure while keeping the target lithium atom fixed.  Six endpoints were identified in which the void space was a local optimum for the lithium atom and in which the relaxation for the rest of the structure was minimal.  These endpoints were then used together with the original \ce{Li_{\alpha}Si} structures as the basis for NEB calculations.  The structures along the resulting NEB path were then passed to the ML model for comparison with the DFT results.

The learning curves are shown in Figure~\ref{fig:TranStatesRMSE}.  The RMSE for the diffusion path structures is less smooth than the RMSE for test folds consisting of the relaxation paths in the amorphous $\text{Li}_{\alpha} \text{Si}$ data. Table \ref{table: energy prediction errors}, second row (diffusion), shows the RMSE and MAE of the per atom energy across all diffusion barrier structures. The RMSE for these structures is about 12.3 meV/atom, which is worse than on the relaxation path test but by less than a factor of two. Nevertheless, reduced accuracy is expected given the extrapolative nature of the task.

However, these errors are not the diffusion barrier errors, which is the quantity of interest. The energies along the diffusion paths are shown in Figure~\ref{fig:TranStates}. The first row plots the absolute energies for both the DFT calculation and the model prediction. The second row shifts the DFT and predicted energy curves to both start at zero, to more easily compare and read off the barriers, which are given in Table~\ref{tab:barriers}. The third row of Figure \ref{fig:TranStates} plots the predicted energy curves as a function of the number of model features $M$, showing the learning rate of the model with respect to this task. The plots indicate that even with a small number of features, for example $M = 21$ or $M = 41$, the energy curve and resulting barrier is qualitatively correct, with additional features serving to refine the curves and better align the total energies. 

Visual inspection of the energy along the diffusion paths shows that much of the error is systematic.  The \ce{Li_{0.2}Si} structures contain 60 atoms, so 12.3 meV/atom corresponds to 0.74 eV in total energy.  If these errors were random, we would expect at least 0.74 eV error in prediction of the diffusion barrier.  However, the curves show that the ML model can successfully distinguish between small-barrier paths and large-barrier paths, and the MAE in barrier prediction is 0.20 eV.  While there is certainly room for improvement, we believe this data shows evidence that the ML model is able to partially capture the physics involved in the diffusion process. 

\subsection{Extrapolate: Larger amorphous systems}
\label{subsec:large_results}

\begin{figure}[t]
\includegraphics[width=0.49\columnwidth]{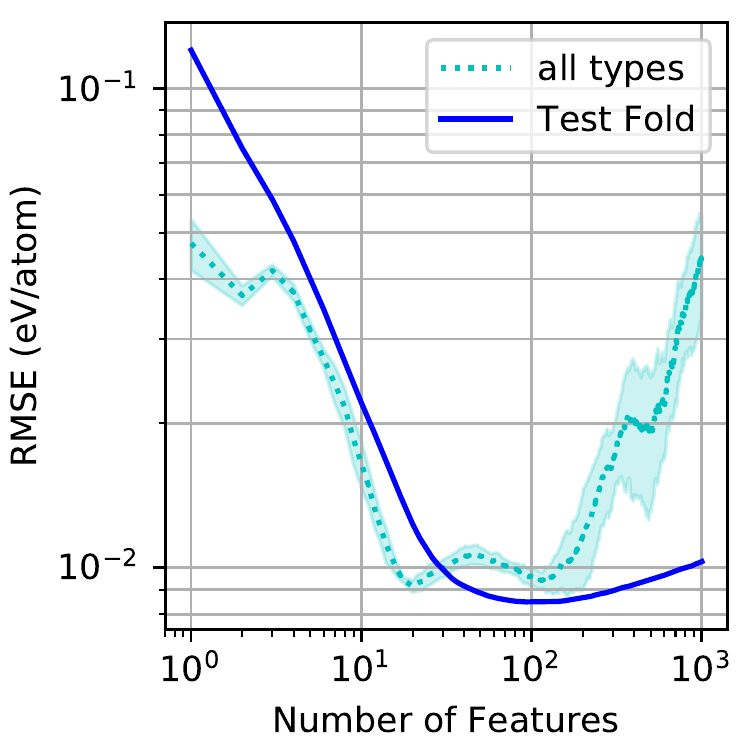}%
\includegraphics[width=0.49\columnwidth]{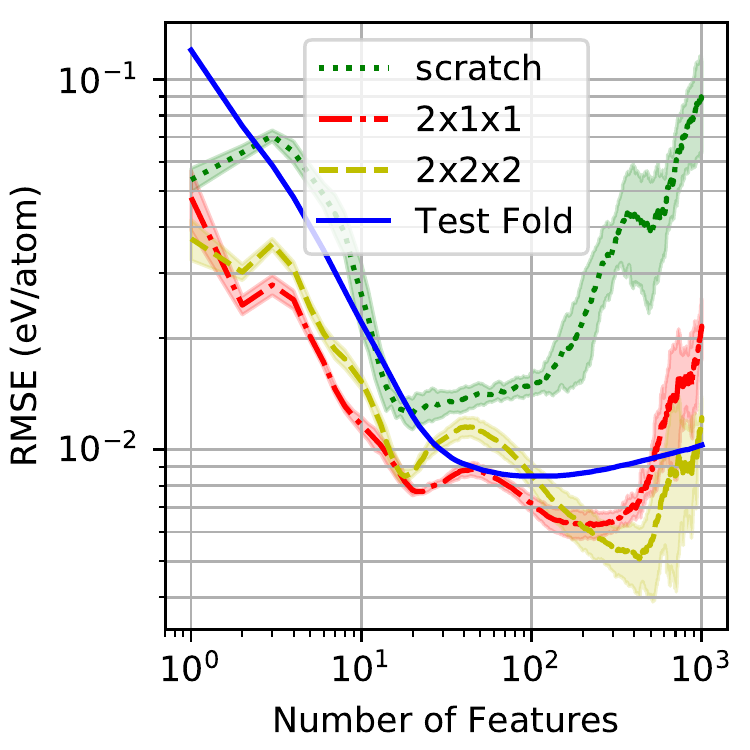}%
\caption{A $\log\text{-}\log$ plot of average of RMSEs of models on the interpolation test set from Section \ref{subsec:dataset_results} and on all types of large states (scratch, 2x1x1, 2x2x2) from Section \ref{subsec:large_results}. Here, y-axis = log(eV/atom), x-axis = log(number of features in models). The curves labeled 2x1x1, 2x2x2, and scratch on right are the RMSE of energy error predictions of the 5 aggregate models separated by test folds. On the left panel, we see that the location of the minimum (i.e., the optimal number of features) for the interpolation test error is similar to the optimal number of features for the extrapolation error on larger states, although model over-fitting is significantly more costly for the larger states' predictions.}
\label{large all} 
\end{figure}

It is desirable for an energy-predictor to generalize to structures in simulation cells with a different size than the training set, so that it can be applied to simulation cells large enough to contain geometries of experimental interest. As system size increases, the computation becomes challenging to carry out with DFT, but the wavelet scattering transform and linear regression scales efficiently with system size (for more details, see Appendix \ref{sec: fast wavelet scattering computations}), and we are thus much less inhibited by large systems. 

\begin{figure}[htbp]
	\centering
	\includegraphics[height=21cm]{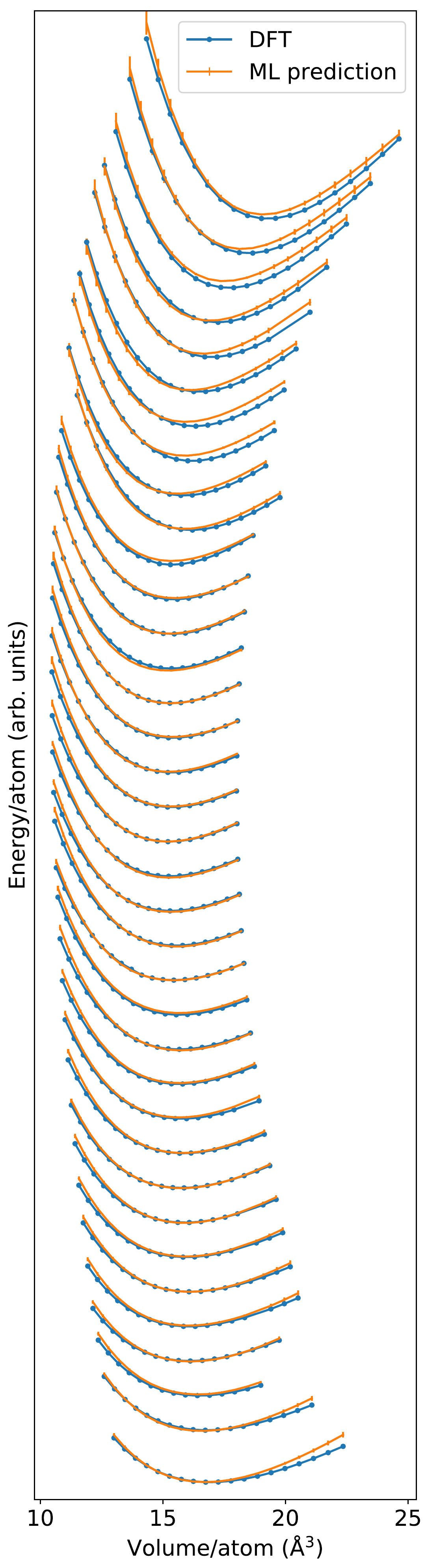}
	\caption{Energy per atom of hydrostatically strained \ce{Li_{\alpha}Si} structures as a function of volume per atom.  Energies are shifted vertically to avoid overlap: $\alpha$ increases down the vertical axis.  Error bars on ML prediction show the standard deviation of predictions of the the 5-fold cross-validated models for each structure. }
	\label{fig:modulus_curves}
\end{figure}

As discussed in Section \ref{sec: data gen for extrapolation}, the data for this task was generated by two different methods: ``from scratch'' and ``tiled.'' The learning curves for each are shown in the right panel of Figure~\ref{large all}. Since our model predicts global energies per atom, it gives the exact same result for a system that is simply periodically duplicated. This suggests that the predictions made when extrapolating to tiled systems that have been perturbed should maintain reasonable accuracy. The figure agrees with this conjecture since the corresponding error lines follow a similar trajectory as the line for the small system test data. The figure also shows that simpler models are favored for the independently relaxed AIMD-generated systems (the ``from scratch'' systems). These systems are less likely than the tiled systems to be similar to examples from the training set. The rapid increase in error on large systems for higher model complexity illustrates the sensitivity of the task to over-fitting. Nevertheless, as depicted in the left panel of Figure \ref{large all},  the optimal number of features for interpolation on amorphous $\text{Li}_{\alpha} \text{Si}$ data is approximately the same as the optimal number of features for energy predictions on the collection of states with larger unit cells. From Table \ref{table: energy prediction errors} (third row, ``large states''), we see that while the prediction errors are higher for the larger systems, it is not an unreasonable increase from errors on the smaller systems.

\subsection{Extrapolate: Bulk Modulus}
\label{subsec:modulus_results}

Elastic properties are another important output of atomistic simulations.  These are typically calculated by applying small strains to the system in question and fitting elastic constants to the energy-versus-strain~\cite{Sholl09}.  This too is an extrapolation task for our model, because uniformly expanded or compressed structures do not appear in the training set.  Testing data for this task was generated based on the lowest-energy structure at each concentration by applying hydrostatic strain, varying the side-length of the simulation box from $-9\%$ to $9\%$.  

\begin{figure}[t]
\includegraphics[width=0.49\columnwidth]{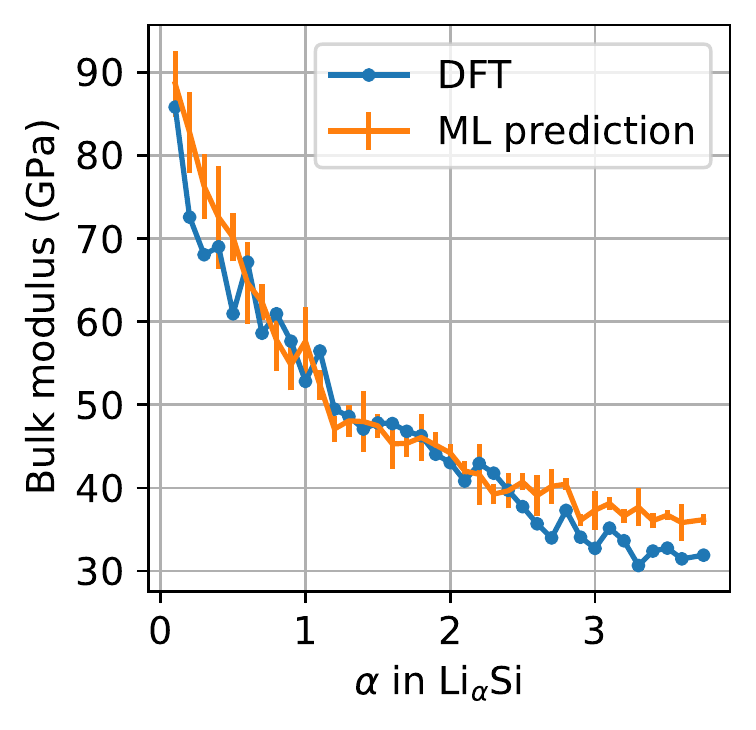}
\includegraphics[width=0.49\columnwidth]{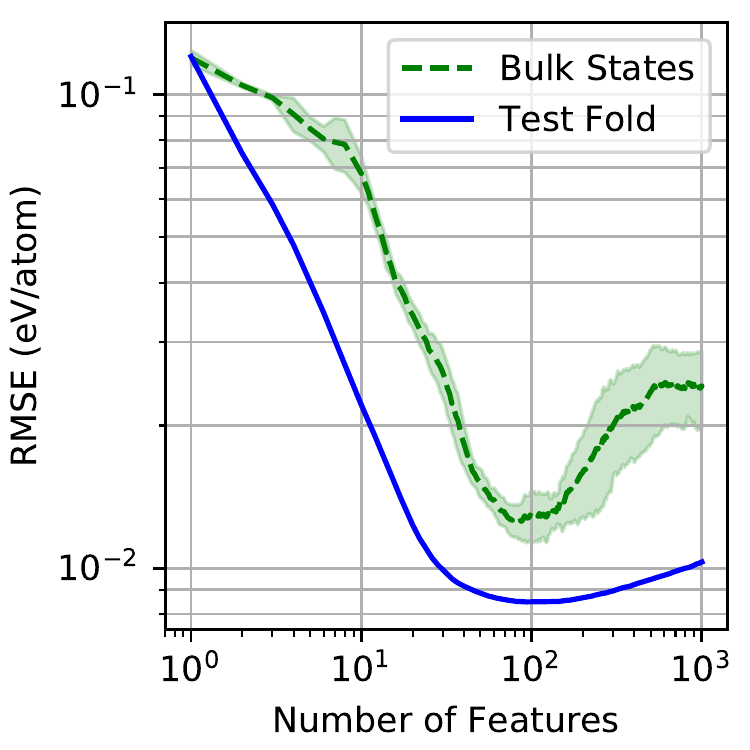}%
\caption{\label{fig:bulk triple resolution test errors} (left) Comparison of DFT-calculated bulk modulus and ML-predicted bulk modulus.  Modulus was calculated through a parabolic fit to points within $\pm$ 4\% strain of the energy minimum.  Error bars on the ML prediction show the standard deviation of fitted modulus across the 5-fold cross-validated models. Averaging across the folds leads to a prediction with MAE of 3.3 GPa compared to the DFT values.  (right) A $\log\text{-}\log$ plot of average of RMSEs of models on the interpolation test set from Section \ref{subsec:dataset_results} and on bulk modulus data from Section \ref{subsec:modulus_results}. Here, y-axis = log(eV/atom), x-axis = log(number of features in models). Green curve is the average of the RMSEs for each fold with error bar given by the standard deviation over the five folds. As for the large states (see Figure \ref{large all}), we see that the location of the minimum (i.e., the optimal number of features) for the interpolation test error is similar to the extrapolation error for the bulk modulus states.}%
\end{figure}

The energy versus volume of the strained structures (Figure~\ref{fig:modulus_curves}) show remarkable agreement between DFT and the ML model.  The RMSE curves shown in Figure~\ref{fig:bulk triple resolution test errors} and the average errors in the last row of Table \ref{table: energy prediction errors} (bulk modulus) are also quite low. The predicted bulk modulus of the structures is shown to decrease as a function of lithium content in Figure~\ref{fig:bulk triple resolution test errors}.  The ML method accurately captures lithiation-induced softening of the silicon. Energy-versus-strain curves along different deformation paths may also be used for the estimation of additional thermodynamic parameters, including Young’s modulus, shear modulus free energies, and heat capacities through the Debye–Gr{\"u}neisen model~\cite{Lu07,Guan19}.

\section{Conclusion}
\label{sec: conclusion}

We have demonstrated a machine-learning model based on atomic orbital wavelet scattering that can achieve an accuracy of 5.52 meV/atom (mean absolute error) in energy on the prediction of amorphous \ce{Li_{\alpha}Si} structures.  We have tested the generalizability of this energy predictor on three extrapolation tasks: diffusion barriers, large systems, and bulk modulus.  As expected based on the nature of regression-based ML, if care is not taken to avoid over-fitting the model performs poorly on these extrapolation tasks.  However, we have shown that a statistically based feature randomization procedure, using the universal wavelet scattering features, can significantly enhance performance on the extrapolation tasks without significant reduction in performance on the interpolative test set.

Though the present work is limited to amorphous \ce{Li_{\alpha}Si}, it provides general lessons for those wishing to apply ML models to new problems in chemical physics.  This is often a daunting task, because ML is generally an interpolative technique.  Before a model can be used, it must be trained on large amounts of data similar to the task at hand.  If the problem is new or challenging to solve by conventional means, the generation of this data can be quite difficult.  Extrapolation from well-known systems may be possible, but off-the-shelf ML models do not extrapolate well.  However, extrapolation performance can be greatly improved by taking a different approach to training the ML model.

Simpler models generalize better.  In our model, ``simplicity'' corresponds to the number of features (wavelet scattering coefficients) used and the fact that these features provide unsupervised descriptions of atomic states, but the concept is general.  Validation sets are often used in machine learning to choose a model complex enough to describe the training data but simple enough to avoid over-fitting. By utilizing randomized feature selection and the aggregation of an ensemble of models (bootstrapping), we obtain a robust and accurate model when applied to the aforementioned extrapolation tasks. From this perspective, typical ML metrics such as testing and validation error are not the only criteria for a ``good'' model.

In order to apply these principles to harder extrapolation tasks and to incorporate a priori uncertainty quantification, it will be necessary to leverage statistical methods that allow one to predict which properties will be difficult for the model, suggesting possibilities for efficient training set expansion to further improve generalizability.  Training set expansion could be automated using ``active learning,'' allowing a model to improve itself based on problems presented to it. The linear regression model over unsupervised nonlinear wavelet scattering features is well positioned for such future work, as it is relatively simple (compared to fully supervised neural networks) to incorporate new data on the fly. 

In future work we will extend the model to include force predictions. As described in \eqref{eqn: scattering energy model} our energy predictions are given as linear combinations of features which are each dependent on the atomic positions. The features are differentiable and we can carry out this differentiation analytically, which opens up fast force computations and fitting the weights of our model to force data. In this case all the methods of Section~\ref{sec: linear regression and model fitting} would still be applicable. This has a computational advantage over features with learned filters which would likely use automatic differentiation. Including forces in the weight learning process could affect the weights that are learned for the energy predictions as each system would have $3N_x$ more points of training data. We expect that this will boost the generalization ability of the model. In this regard, training on forces will act as a regularizer for energy predictions.

\section*{Data Availability Statement}
The data that support the findings of this study are available from the corresponding author upon reasonable request.

\begin{acknowledgments}
This work was supported in part by the Defense Advanced Research Projects Agency [Young Faculty Award D16AP00117 to M.H., supporting P.S., X.B., J.L., and K.J.K]; the Alfred P. Sloan Foundation [Sloan Fellowship FG-2016-6607 to M.H.]; the National Science Foundation [grant 1832808 to Y.Q., grant 1620216 to M.H., CAREER award 1845856 to M.H], and the Michigan State University Foundation [Strategic Partnership Grant to Y.Q.]. This work used computational resources provided  by  the  Institute for Cyber-Enabled Research at Michigan State University. X.B. also acknowledges support from the Institute for Pure and Applied Mathematics at UCLA, where he was in residence during fall 2019.
\end{acknowledgments}

\appendix

\section{Channel definitions}
\label{sec: channel def}

In this appendix we give precise definitions of the five input channels:
\begin{itemize}[itemsep=0pt, topsep=0pt]
	\item Lithium channel: $f(Z_k) = Z_k$ if $Z_k = 3$ and $f(Z_k) = 0$ otherwise
	\item Silicon channel: $f(Z_k) = Z_k$ if $Z_k = 14$ and $f(Z_k) = 0$ otherwise
	\item Valence channel: $f(Z_k) = $ \# of valence electrons
	\item Ionic channel: $f(Z_k) = $ \# of core electrons
	\item Kinetic channel: $f(Z_k) = \sqrt{Z_k}$
\end{itemize}

\section{Fast wavelet scattering computations}
\label{sec: fast wavelet scattering computations}

In this appendix we describe how to efficiently compute the wavelet scattering features described in Section \ref{subsec:ML_algorithm}. 

In practice all computations are carried out over a discrete sampling of the unit cell $\Qc_x$. Such a sampling is a three-dimensional grid $\Gc_x \subset \Qc_x$ with $L_x$ grid points along each dimension. Due to the multiscale sizes of the wavelet filters $\psi_{j, n, \ell}^m$ a direct computation of the circular convolution \eqref{eqn: circular convolution} over the grid $\Gc_x$ will require $O(L_x^6)$ floating point operations. This computational cost can be significantly reduced by carrying out these computations in frequency space. 

Recalling \eqref{eqn: regular convolution}, the Fourier transform of $\rho_x^f \ast \psi_{j, n, \ell}^m$ is:
\begin{equation} \label{eqn: fourier transform wavelet coeffs}
    \Fc [\rho_x^f \ast \psi_{j, n, \ell}^m] (\omega) = \widehat{\psi}_{n, \ell}^m (2^j \omega) \sum_{k = 1}^{N_x} f(Z_k) e^{-i \omega \cdot R_k} \, ,
\end{equation}
where $\Fc [h] (\omega) = \widehat{h}(\omega)$ is the Fourier transform of the function $h \in \Lb^1 (\R^3)$. The Fourier transform of $\psi_{n, \ell}^m$ can be computed analytically:
\begin{equation*}
    \widehat{\psi}_{n, \ell}^m (\omega) = (-i)^{\ell} \sqrt{\frac{4\pi}{2 \ell + 1}} |\omega|^{2(n-1) - \ell} e^{-\beta^2 |\omega|^2 / 2} Y_{\ell}^m (\omega / |\omega|) \, .
\end{equation*}
Therefore \eqref{eqn: fourier transform wavelet coeffs} can be evaluated directly for any $\omega \in \R^3$. We do so in a box $[-\omega_x, \omega_x)^3$, where 
\begin{equation*}
    \omega_x = \frac{\pi}{\Delta_x} \quad \text{and} \quad \Delta_x = \frac{s_x}{L_x} \, .
\end{equation*}
The maximum frequency $\omega_x$ is chosen so that the essential support of $\widehat{\psi}_{n, \ell}^m$ is contained within $[-\omega_x, \omega_x)^3$, which in turn determines the number of grid points $L_x$ along each side length of the unit cell $\Qc_x$. This maximum frequency depends on the wavelet width $\beta$ and (weakly) on the $(n, \ell)$ parameters. Evaluations within the box $[-\omega_x, \omega_x)^3$ are restricted to a grid $\Omega_x \subset [-\omega_x, \omega_x)^3$ with grid spacing $2\pi / s_x$, which yields $L_x$ frequency grid points. In particular we compute, via direct numerical evaluation, a tensor $\Psi_{j, n, \ell}^m \in \C^{L_x} \times \C^{L_x} \times \C^{L_x}$ defined as
\begin{equation} \label{eqn: discrete frequency evaluation}
    \Psi_{j, n, \ell}^m = \Fc [\rho_x^f \ast \psi_{j, n, \ell}^m] \Big|_{\Omega_x} \, .
\end{equation}
Due to the discretization in \eqref{eqn: discrete frequency evaluation}, taking the inverse fast Fourier transform (iFFT) of $\Psi_{j, n, \ell}^m$ recovers the circular convolution $\rho_x^f \circledast \psi_{j, n, \ell}^m$ evaluated on the spatial grid $\Gc_x$:
\begin{equation} \label{eqn: discrete circular convolution}
    \mathrm{iFFT} (\Psi_{j, n, \ell}^m) = \rho_x^f \circledast \psi_{j, n, \ell}^m \Big|_{\Gc_x} \, .
\end{equation}
The direct computation of $\Psi_{j, n, \ell}^m$ requires $C N_x L_x^3$ floating point operations, whereas the iFFT calculation requires $C L_x^3 \log L_x$ floating point operations. Therefore the total cost is reduced to $O ((N_x + \log L_x) L_x^3)$.

First order wavelet scattering features are estimated by applying the pointwise nonlinear operator $\sigma$ to \eqref{eqn: discrete circular convolution} and estimating the $\Lb^q (\Qc_x)$ integrals with a Riemann sum approximation. Second order wavelet scattering features are computed by taking the fast Fourier transform (FFT) of $\sigma (\rho_x^f \circledast \psi_{j, n, \ell}) \big|_{\Gc_x}$ and computing the second circular wavelet convolution via frequency multiplication with a direct evaluation of $\widehat{\psi}_{n_2, \ell_2}^{m_2} (2^{j_2} \omega)$ on the grid $\omega \in \Omega_x$, followed by another iFFT, application of $\sigma$, and Riemann sum. The cost of each second order feature, given that \eqref{eqn: discrete circular convolution} must already be computed for the first order features, is $O(L_x^3 \log L_x)$. 

\section{Alternative Models Comparison}
\label{sec: alternative models}

In this appendix we describe two models similar to the one used in the main body of this text and compare the results.

The model used in the main body (hereafter referred to as the full model) has numerical results on the test set summarized in Table~\ref{table: energy prediction errors} and the training method is described in Section~\ref{sec: linear regression and model fitting}. The results of two alternative models on the various tasks of this work are listed in Tables~\ref{table: 0,1-order energy prediction errors}~and~\ref{table: traditionally-trained energy prediction errors}. The test folds of relaxation paths and \ce{Li_\alpha Si} states of the diffusion, large states, and bulk modulus states are identical in all three model comparisons.

\begin{table}[t]
\begin{tabular}{|l|l|l|}
\hline \hline
& RMSE (meV/atom) & MAE (meV/atom) \\ \hline
Relaxation paths & 8.04 $\pm$ 0.59 & 5.99 $\pm$ 0.39 \\ 
Diffusion & 11.8 $\pm$ 0.48 & 9.51 $\pm$ 0.95 \\ 
Large states & 14.0 $\pm$ 0.68 & 10.2 $\pm$ 0.42 \\ 
Bulk modulus & 39.5 $\pm$ 4.82 & 25.1 $\pm$ 2.92 \\ \hline
\hline
\end{tabular}
\caption{Numerical results for ML predictions with only zero and first order features (compared to zero, first, and second in Table~\ref{table: energy prediction errors}) on the test data from the amorphous dataset and the three extrapolation tasks from the model trained only on the amorphous data.}
\label{table: 0,1-order energy prediction errors}
\end{table}

The first alternative model (hereafter the 0-1 model) is trained identically to the full model with five test folds (the test folds are identical for both models) and 100 randomly selected sets of relaxation strings (with replacement) for training, but with only zero and first order wavelet scattering features available for selection in training. This results in a total of 321 features (with bias) to select from compared to 3741 in the full model. Note that at each step of the greedy OLS training the best feature is chosen from $\sqrt{321} \approx 17$ features that are randomly selected from the remaining unselected features compared to $\sqrt{3741} \approx 61$ in the full model. The model size $M^{\star}$ averaged over all 500 permutations of the training data is 121 in the full model and 108 for the 0-1 model, with standard deviations of 64 and 38, respectively. The numerical results for the 0-1 model are listed in Table~\ref{table: 0,1-order energy prediction errors}. The performance is comparable on the relaxation paths. On the diffusion states the 0-1 model has slightly better performance in RMSE and MAE, but the standard deviation in MAE across the five folds is nearly double the full model. Furthermore, inspection of the barriers computed by the 0-1 model reveals that they are in fact slightly worse than the full model. The performance of the 0-1 model is significantly worse than the full model on the large and bulk states, again with a large spread in errors. This indicates that we get a statistically significant benefit by including second order features in the models. 

\begin{table}[t]
\begin{tabular}{|l|l|l|}
\hline \hline
& RMSE (meV/atom) & MAE (meV/atom) \\ \hline
Relaxation paths & 7.50 $\pm$ 0.39 & 5.64 $\pm$ 0.28 \\ 
Diffusion & 11.6 $\pm$ 1.01 & 11.0 $\pm$ 1.03 \\ 
Large states & 9.78 $\pm$ 1.98 & 6.60 $\pm$ 0.81 \\ 
Bulk modulus & 16.6 $\pm$ 4.91 & 11.5 $\pm$ 3.55 \\ \hline
\hline
\end{tabular}
\caption{Numerical results for ML predictions with the non-randomized model. The models are trained without random feature selection at each step of the greedy OLS algorithm, i.e., at each step all features are available for selection.}
\label{table: traditionally-trained energy prediction errors}
\end{table}

The second alternative model (hereafter the non-randomized model) has the same features available as the full model and the same five test folds as the prior two models. The training set is randomly partitioned into four equally sized sets (selection by relaxation strings) with a model trained for each selection of a set as validation and the remaining three for training (i.e., nested five-fold cross validation, as in~ \citet{hansen:quantumChemML2013}). This results in four trainings for each test fold for a total of 20 models trained compared to the 500 trainings (five test sets with 100 training/validation splits) of the full model. This non-randomized procedure ensures uniform representation of the strings in the training, validation, and testing folds. During training of the non-randomized model the OLS algorithm seeks the next best feature at each step from all remaining features rather than randomly selecting a subset of features to choose from as in the prior two models. The average value of $M^{\star}$ is 153 for the non-randomized model with standard deviation of 82 across the 20 trainings. The performance of this model is similar to the full model on relaxation paths but with significantly larger spread of the errors between models on the diffusion, large, and bulk states. Thus the chance of a catastrophic error is higher. Furthermore, the RMSE and MAE are significantly larger on the bulk states. This indicates that the model over-fit the training data and did not generalize as well to the extrapolation tasks. Randomized training in the full model appears to mitigate the possibility of over-fitting.

\section*{Supplementary Material}
\label{sec: supplementary material}

\subsubsection*{Validation of amorphous structures}

\begin{figure}
	\centering
	\includegraphics[width=0.9\columnwidth]{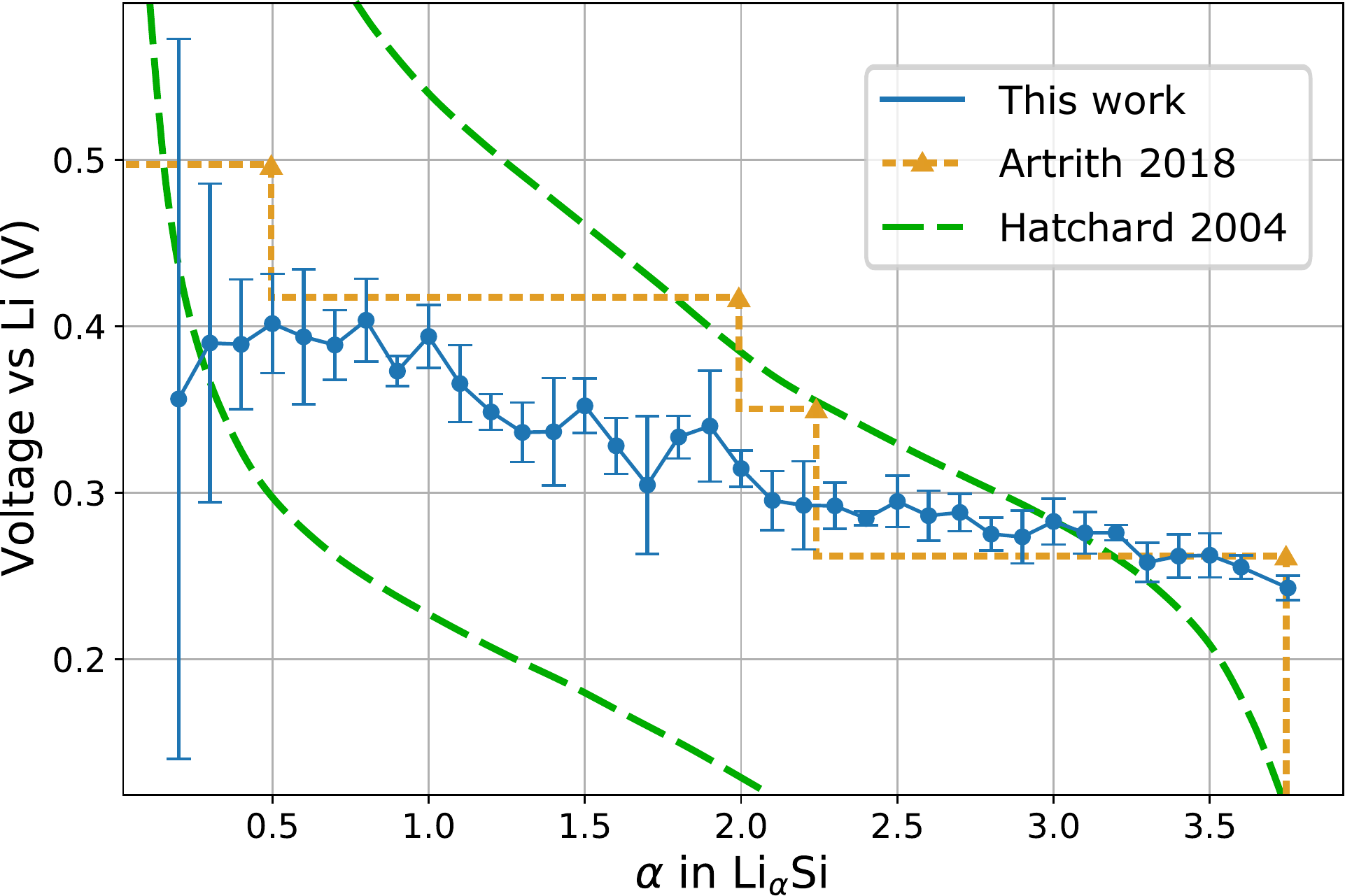}
	\caption{Voltage versus Li/Li$^+$ of fully relaxed amorphous Li$_\alpha$Si structures, plotted as a function of $\alpha$ (blue circles and solid lines).  Voltages are averaged over the 10 fully relaxed structures at each $\alpha$ in our data set, and error bars show the standard deviation.  Shown for comparison are experimental results from Ref.~\citenum{Hatchard04} (green dashed lines) and the results from the genetic algorithm ensemble of Ref.~\onlinecite{Artrith:LiSiML2018} (orange dashed lines and triangles). }
	\label{fig:voltage}
\end{figure}

In order to validate the amorphous structures generated by our method (See Section II.A. of the main text), we calculate the average voltage of lithiation from Li$_{0.1}$Si to Li$_\alpha$Si versus Li/Li$^+$ through the equation
\begin{equation}
	V_\alpha = -\frac{E[\text{Li}_\alpha\text{Si}] - E[\text{Li}_{0.1}\text{Si}] - (\alpha-0.1)E[\text{Li}]}{(\alpha-0.1)e}
\end{equation}
We compare the result with experimental lithiation and delithiation voltages from~ \citet{Hatchard04} and the genetic algorithm sampling from~ \citet{Artrith:LiSiML2018} in Figure~\ref{fig:voltage}, finding good agreement.

\begin{figure*}
	\centering
	\includegraphics[width=0.75\textwidth]{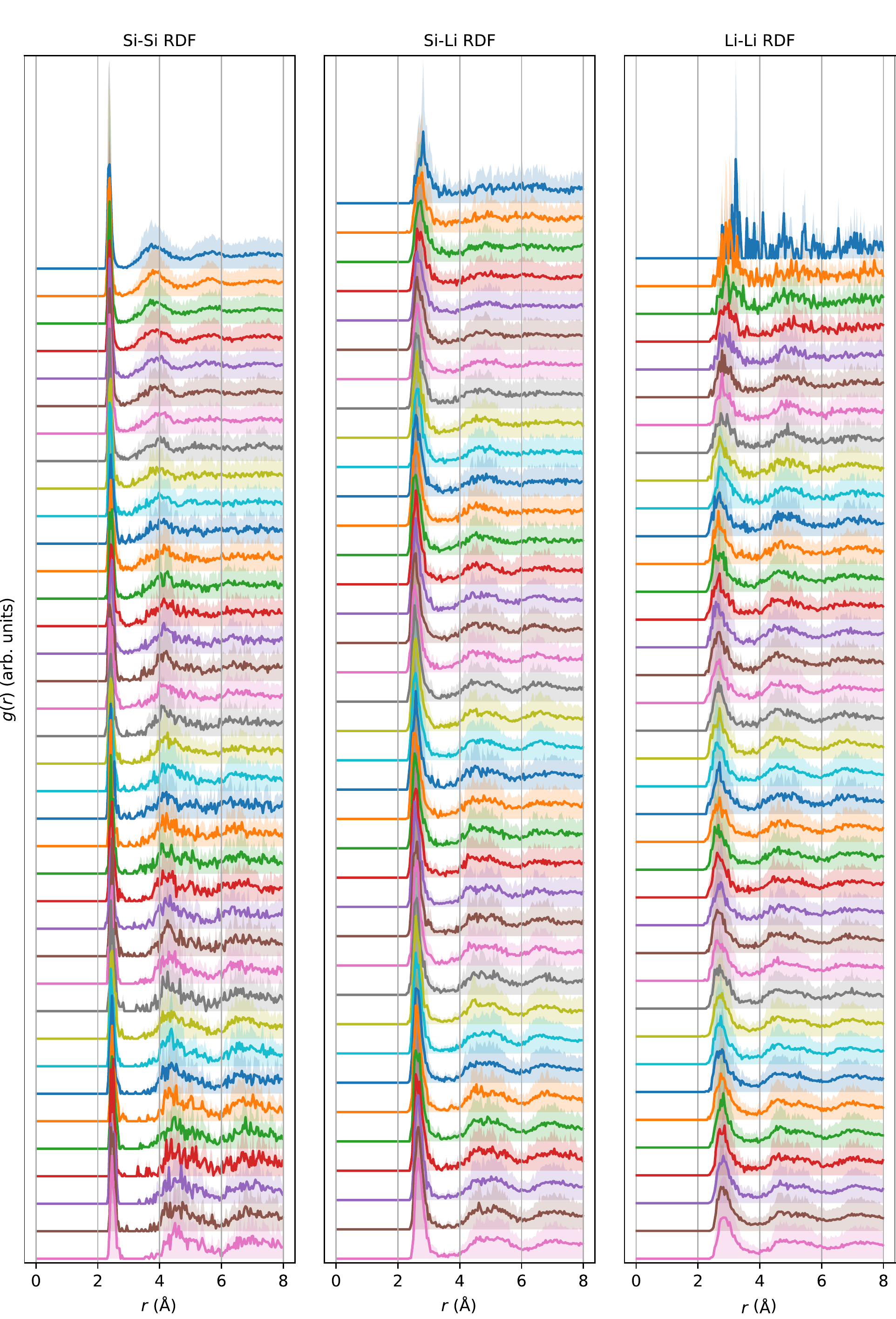}
	\caption{Radial distribution functions of Si-Si, Si-Li, and Li-Li pairs in fully relaxed amorphous Li$_\alpha$Si structures in our data set.  The lithium fraction $\alpha$ increases from 0.1 to 3.75 going down the vertical axis.  RDFs are averaged over the 10 fully relaxed structures at each $\alpha$ in our data set, and shaded regions show the standard deviation.  The first peak height is comparable with literature reports~\cite{Artrith:LiSiML2018}, and further peaks show expected trends with significant variation, showing a healthy variety of atomic environments for ML training. }
	\label{fig:rdf}
\end{figure*}

We also calculate the radial distribution function (RDF) for the Si-Si, Si-Li, and Li-Li pairs in the amorphous structures.  These are plotted in Figure~\ref{fig:rdf}, and show good agreement with similar results in the literature~\cite{Artrith:LiSiML2018}.

\bibliography{MainBib.bib}

\end{document}